\documentclass[aps, pra, amsmath, amssymb, reprint]{revtex4-1}

\usepackage[]{graphicx}
\usepackage[]{hyperref}
\usepackage{bm}

\newcommand{\no}{\nonumber}
\newcommand{\drm}{\text{d}}
\newcommand{\im}{\text{i}}
\newcommand{\grad}{\nabla}
\newcommand{\e}{\text{e}}
\newcommand{\del}{\partial}

\newcommand{\rbm}{\bm{r}}
\newcommand{\kbm}{\bm{k}}
\newcommand{\qbm}{\bm{q}}
\newcommand{\Abm}{\bm{A}}

\newcommand{\Tc}{T_\text{c}}
\newcommand{\TcR}{T_\text{cR}}
\newcommand{\EF}{E_\text{F}}

\newcommand{\Tr}{\text{Tr} \,}
\newcommand{\Ntot}{N_\text{tot}}
\newcommand{\ntot}{n_\text{tot}}
\newcommand{\cbar}{\overline{c}}
\newcommand{\Pbar}{\overline{P}}
\newcommand{\abar}{\overline{a}}
\newcommand{\SGL}{\mathcal{S}_\text{GL}}
\newcommand{\Seff}{\mathcal{S}_\text{eff}}
\newcommand{\xiGL}{\xi_0}
\newcommand{\ftri}{f_\text{tri}}
\newcommand{\fopt}{f_\text{opt}}
\newcommand{\qc}{q_\text{c}}
\newcommand{\aR}{a_\text{R}}

\newcommand{\xiR}{\xi_\text{0R}}
\newcommand{\epsilonR}{\epsilon_\text{R}}
\newcommand{\hR}{h_\text{R}}
\newcommand{\Bctwo}{B_\text{c2}}
\newcommand{\BctwoR}{B_\text{c2R}}
\newcommand{\Mdia}{M_\text{dia}}
\newcommand{\chidia}{\chi_\text{dia}}

\begin{document}
\title{Enhanced superconducting-fluctuation effects on thermodynamic properties in BCS-BEC-crossover regime}
\author{Kyosuke Adachi}
\author{Ryusuke Ikeda}
\affiliation{Department of Physics, Kyoto University, Kyoto 606-8502, Japan}
\date{\today}
\begin{abstract}
Effects of superconducting fluctuation (SCF) on thermodynamic properties of electron systems in the so-called BCS-BEC-crossover regime are studied.
As the attractive interaction between electrons becomes stronger upon approaching the BCS-BEC-crossover regime, importance of the mode coupling between SCF drastically increases.
The enhanced mode coupling leads to lowering of both the zero-field superconducting critical temperature and the depairing field $\Bctwo (T)$.
Consequently, SCF-induced contributions to the specific heat and the diamagnetic susceptibility can seemingly exceed the corresponding values in the Gaussian approximation.
We discuss relevance of the present results to the anomalous SCF-induced diamagnetic response observed in the iron selenide (FeSe).
\end{abstract}
\maketitle

\section{Introduction}
\label{Sec:Introduction}

Several unique electronic properties have been observed in the iron selenide (FeSe)~\cite{Hsu_Luo_2008}, one of the iron-based superconductors.
According to the recent quantum-oscillation measurement~\cite{Terashima_Kikugawa_2014} and scanning tunneling spectroscopy~\cite{Kasahara_Watashige_2014}, the Fermi energy of FeSe is comparable in magnitude to the superconducting gap.
This fact suggests an intriguing possibility that electrons in FeSe effectively interact with a much stronger attractive interaction than those in conventional superconductors.
In other words, electrons in FeSe can be in the so-called BCS-BEC-crossover regime~\cite{Randeria_Taylor_2014}, where it is theoretically proposed that preformed Cooper pairs and a pseudogap in the one-particle density of states emerge around a pairing onset temperature $T^*$ far above the superconducting critical temperature $\Tc$.
In fact, the NMR relaxation rate and transport coefficients in this material show unusual changes when the temperature gets across about $2 \Tc$, which is a candidate of $T^*$~\cite{Kasahara_Yamashita_2016}.

Moreover, effects of superconducting fluctuation (SCF) near $\Tc$ are also peculiar in FeSe~\cite{Kasahara_Yamashita_2016}.
The observed SCF-induced diamagnetic susceptibility in low magnetic fields is anomalously large compared to the theoretically predicted value in the Gaussian approximation.
Also the temperature-dependent magnetization curves in high magnetic fields show a SCF-induced crossing behavior~\cite{Kasahara_Yamashita_2016}.
This crossing behavior has been observed mainly in strongly two-dimensional (2D) systems such as the BSCCO compounds of the high-$\Tc$ cuprates~\cite{Li_Suenaga_1992} and has been theoretically supported in a 2D model~\cite{Tesanovic_Xing_1992}.
In contrast, FeSe is 3D-like judging from the fact that the size of the coherence length $\xi_{0, c}$ ($\sim 1.3\text{nm}$~\cite{Terashima_Kikugawa_2014}) along the $c$-axis is longer than $s / \sqrt{2}$, where $s$ ($\sim 0.55\text{nm}$~\cite{Kasahara_Watashige_2014}) is the interlayer spacing or the $c$-axis length.
Hence the crossing behavior of the magnetization curves observed in FeSe~\cite{Kasahara_Yamashita_2016} is an unexpected event.

In the BCS-BEC-crossover regime, $\Tc$ is high, and the coherence length is short compared with those in the weak-coupling BCS regime, so that the critical region is enlarged~\cite{Debelhoir_Dupuis_2016}.
Thus it is natural that the SCF should be enhanced when the system enters the BCS-BEC-crossover regime.
However, it is non-trivial whether the SCF in the BCS-BEC-crossover regime affects thermodynamics in the same manner as that in the BCS regime.
Further, a lot of experiments on SCF in FeSe have been performed in finite magnetic fields.
Therefore, it is an important subject to theoretically investigate whether the idea that FeSe is in the BCS-BEC-crossover regime can explain the observed strange behaviors such as the enhanced SCF effect on the diamagnetic response.

On the basis of these backgrounds, we consider SCF effects on thermodynamics of a system under magnetic fields in the BCS-BEC-crossover regime and elucidate some features which are substantially different from those in the BCS regime.
This paper is organized as follows.
In Sec.~\ref{Sec:Method}, we explain the theoretical model and method used to estimate the SCF-induced specific heat and diamagnetism.
In Sec.~\ref{Sec:Results}, we present our calculations of thermodynamic quantities in addition to some preliminary results on the critical temperature and coherence length.
In Sec.~\ref{Sec:Discussion}, we discuss relevance of our results to the anomalous SCF-induced phenomena observed in FeSe under magnetic fields and add some remarks.
Finally in Sec.~\ref{Sec:Conclusion}, we state our conclusion.

\section{Method}
\label{Sec:Method}

\subsection{Model}
\label{SubSec:Model}

To consider the SCF-induced thermodynamic quantities such as the specific heat and the diamagnetic response, we start with a simple Hamiltonian of an isotropic three-dimensional system with a separable attractive interaction:
\begin{eqnarray}
H &=& H_0 + H_\text{int} \no \\
&=& \sum_{\kbm, \sigma} \frac{k^2}{2 m} c_{\kbm \sigma}^\dag c_{\kbm \sigma} \no \\
&& - \frac{U}{V} \sum_{\qbm, \kbm, \kbm'} \varphi_k \varphi_{k'} c_{\qbm / 2 + \kbm \uparrow}^\dag c_{\qbm / 2 - \kbm \downarrow}^\dag c_{\qbm / 2 - \kbm' \downarrow} c_{\qbm / 2 + \kbm' \uparrow}. \no \\
\label{Eq:Hamiltonian}
\end{eqnarray}
Here, $m$ is the mass of particles, $U (> 0)$ is the interaction strength, $V$ ($= L_x L_y L_z$) is the total volume of the system, and $c_{\kbm \sigma}^{(\dag)}$ is the annihilation (creation) operator of a particle with spin $\sigma$ and momentum $\kbm$.
The interaction form factor $\varphi_k$ is introduced as~\cite{Nozieres_Schmitt-Rink_1985}
\begin{equation}
\varphi_k = \frac{1}{\sqrt{\displaystyle 1 + (k / k_0)^2}},
\label{Eq:FormFactor}
\end{equation}
where $k_0$ is an effective momentum cutoff.

We assume a system with the total number (density) of particles fixed to $\Ntot$ ($\ntot$) and consider a grand-canonical ensemble specified by the temperature $T$ ($= \beta^{-1}$) and the chemical potential $\mu$.
Effects of an applied magnetic field effect is taken into account afterwards (see Sec.~\ref{SubSec:GL_action_describing_the_zero-field_SCF}).
In a two-particle system described with Eq.~(\ref{Eq:Hamiltonian}), a two-particle bound state can appear~\cite{Maly_Janko_1999} when $U$ is strong enough to exceed a threshold value [$U_0 = 4 \pi / (m k_0)$].
Thus we expect that a many-body system described with Eq.~(\ref{Eq:Hamiltonian}) will be in the BCS-BEC-crossover regime when $U$ is close to $U_0$.

Though the model described with Eq.~(\ref{Eq:Hamiltonian}) is clearly too simple to describe the electron states in FeSe, we believe that we can sufficiently study with this model the generic nature of the SCF effects on thermodynamics in the BCS-BEC-crossover regime.

\subsection{Shift of chemical potential}
\label{SubSec:Shift_of_chemical_potential}

In the BCS-BEC-crossover regime, a strong attractive interaction causes a decrease in the chemical potential from the Fermi energy defined as $(3 \pi^2 \ntot)^{2 / 3} / (2 m)$ ($= \EF$).
To determine the chemical potential, we calculate the thermodynamic potential in zero field following the standard method developed by Nozi{\`e}res and Schmitt-Rink~\cite{Nozieres_Schmitt-Rink_1985}.
First, the thermodynamic potential $\Omega$ is calculated within the ladder approximation, which is equivalent to the Gaussian approximation, as
\begin{equation}
\Omega = \Omega_0 + T \sum_{\qbm, m} \e^{+ \im \omega_m 0} \ln \left[ 1 - U K_{\qbm} (\im \omega_m) \right],
\end{equation}
where $\Omega_0$ [$= - T \ln \Tr \exp (- \beta H_0 + \beta \mu \Ntot)$] is the contribution from the kinetic energy of the electrons, $\omega_m$ ($ =2 \pi m T$) ($m = 0, \pm 1, \pm 2, \cdots$) is the Bosonic Matsubara frequency.
The bare superconducting susceptibility $K_{\qbm} (\im \omega_m)$ is defined as
\begin{eqnarray}
&& K_{\qbm} (\im \omega_m) \no \\
&=& \frac{T}{V} \sum_{\kbm, n} {\varphi_k}^2 G_{\qbm / 2 + \kbm \uparrow} (\im \varepsilon_n + \im \omega_m) G_{\qbm / 2 - \kbm \downarrow} (- \im \varepsilon_n) \no \\
&=& \frac{1}{2V} \sum_{\kbm} {\varphi_k}^2 \frac{\tanh (\beta \xi_{\qbm / 2 + \kbm} / 2) + \tanh (\beta \xi_{\qbm / 2 - \kbm} / 2)}{\xi_{\qbm / 2 + \kbm} + \xi_{\qbm / 2 - \kbm} - \im \omega_m}, \no \\
\end{eqnarray}
where $\xi_{\kbm} = \xi_{k}= k^2 / (2 m) - \mu$ and $G_{\kbm \sigma} (\im \varepsilon_n)$ is the bare electron propagator defined as
\begin{equation}
G_{\kbm \sigma} (\im \varepsilon_n) = \frac{1}{\xi_{\kbm} - \im \varepsilon_n}.
\end{equation}
Here $\varepsilon_n$ [$= 2 \pi (n + 1/2) T$] ($n = 0, \pm 1, \pm 2, \cdots$) is the Fermionic Matsubara frequency.
Next, by differentiating $\Omega$ with respect to $\mu$, we get the total number density
\begin{eqnarray}
&&\ntot \no\\
=&& - \frac{1}{V} \frac{\del \Omega}{\del \mu} \no \\
=&& \frac{1}{V} \sum_{\kbm, \sigma} \frac{1}{\e^{\beta \xi_{\kbm}} + 1} + \frac{T}{V} \sum_{\qbm, m} \e^{+ \im \omega_m 0} \frac{U \del_\mu K_{\qbm} (\im \omega_m)}{1 - U K_{\qbm} (\im \omega_m)}.
\label{Eq:NumberEquation}
\end{eqnarray}
Further, in the Gaussian approximation, the superconducting critical temperature ($\Tc$) is determined by
\begin{equation}
1 - U K_{\bm{0}} (0) = 0.
\label{Eq:TcEquation}
\end{equation}
From Eqs.~(\ref{Eq:NumberEquation}) and (\ref{Eq:TcEquation}), we obtain both $\Tc$ and $\mu (T = \Tc)$ for the fixed $\ntot$.

\subsection{GL action describing zero-field SCF}
\label{SubSec:GL_action_describing_the_zero-field_SCF}

In this subsection, we derive the Ginzburg-Landau (GL) action to describe the SCF effects on thermodynamics near $\Tc$.
First, we rewrite the grand-canonical partition function $Z$ in the functional-integral form~\cite{Melo_Randeria_1993}:
\begin{equation}
Z = \Tr \e^{- \beta H + \beta \mu \Ntot} = \int \left( \prod_{\kbm, \sigma, n} \drm \cbar_{\kbm \sigma n}  \drm c_{\kbm \sigma n} \right) \e^{-\mathcal{S}},
\end{equation}
where the action $\mathcal{S}$ is defined as
\begin{eqnarray}
\mathcal{S} &=& \mathcal{S}_0 + \mathcal{S}_\text{int} \no \\
&=& \beta \left[ \sum_{\kbm, \sigma, n} \left( - \im \varepsilon_n + \xi_{\kbm} \right) \cbar_{\kbm \sigma n} c_{\kbm \sigma n} - \frac{U}{V} \sum_{\qbm, m} \Pbar_{\qbm m} P_{\qbm m} \right].  \no \\
\label{Eq:Action}
\end{eqnarray}
Here, the dimensionless Fermionic fields $\{ c_{\kbm \sigma n}, \cbar_{\kbm \sigma n} \}$ are the Grassmann numbers, and
\begin{equation}
\left\{
\begin{array}{l}
\displaystyle P_{\qbm m} = \sum_{\kbm, n} \varphi_k  c_{\qbm / 2 - \kbm \downarrow -n - 1} c_{\qbm / 2 + \kbm \uparrow n + m} \vspace{5pt} \\
\displaystyle \Pbar_{\qbm m} = \sum_{\kbm, n} \varphi_k \cbar_{\qbm / 2 + \kbm \uparrow n + m} \cbar_{\qbm / 2 - \kbm \downarrow -n - 1}
\end{array}
\right.
\end{equation}
Next, we use the Hubbard-Stratonovich transformation~\cite{Melo_Randeria_1993}.
By introducing the dimensionless Bosonic order-parameter fields $\{ a_{\qbm m}, \abar_{\qbm m} \}$ corresponding to the superconducting order-parameter fields, we rewrite the interaction part of the action $\mathcal{S}_\text{int}$ as
\begin{eqnarray}
\e^{-\mathcal{S}_\text{int}} &=& \int \left( \prod_{\qbm, m} \frac{\drm \abar_{\qbm m} \drm a_{\qbm m}}{\pi} \right) \e^{-\sum_{\qbm, m} \abar_{\qbm m} a_{\qbm m}} \no \\
&& \times \e^{\sqrt{\beta U / V} \sum_{\qbm, m} (\abar_{\qbm m} P_{\qbm m} + a_{\qbm m} \Pbar_{\qbm m})}.
\end{eqnarray}
Then, by integrating with respect to $\{ c_{\kbm \sigma n}, \cbar_{\kbm \sigma n} \}$, we obtain
\begin{equation}
Z = Z_0 Z_\text{eff},
\end{equation}
where $Z_0$ ($= e^{-\beta \Omega_0}$) is the non-interacting part of $Z$, and
\begin{equation}
Z_\text{eff} = \int \left( \prod_{\qbm, m} \frac{\drm \abar_{\qbm m} \drm a_{\qbm m}}{\pi} \right) \e^{- \Seff}.
\label{Eq:Zeff}
\end{equation}
Here, $\Seff$ is the effective action describing the SCF, formally expressed as
\begin{equation}
\e^{- \Seff} = \e^{-\sum_{\qbm, m} \abar_{\qbm m} a_{\qbm m}} \langle \e^{\sqrt{\beta U / V} \sum_{\qbm, m} (\abar_{\qbm m} P_{\qbm m} + a_{\qbm m} \Pbar_{\qbm m})} \rangle_0,
\label{Eq:Seff}
\end{equation}
where $\langle \cdots \rangle_0$ denotes the grand-canonical average with respect to the non-interacting part $\mathcal{S}_0$.

We can reproduce Eqs.~(\ref{Eq:NumberEquation}) and (\ref{Eq:TcEquation}) by expanding $\Seff$ up to the second order in $\{ a_{\qbm m}, \abar_{\qbm m} \}$, \textit{i.e.}, using the Gaussian approximation.
However, since the critical region is strongly enhanced in the BCS-BEC-crossover regime~\cite{Debelhoir_Dupuis_2016}, we need to go beyond the Gaussian approximation, \textit{i.e.}, the mode coupling between the SCF has to be incorporated to examine the critical behavior.

To treat the mode coupling, we expand $\Seff$ up to the fourth order in $\{ a_{\qbm m}, \abar_{\qbm m} \}$.
In addition, we neglect the quantum fluctuation (\textit{i.e.}, $a_{\qbm m}$ and $\abar_{\qbm m}$ with $m \neq 0$) and use the gradient expansion.
This results in replacing $\Seff$ in Eq.~(\ref{Eq:Zeff}) with the GL action $\SGL$:
\begin{equation}
\SGL = \int \drm^3 \rbm \left\{ a \left[ \epsilon |\psi|^2 + {\xiGL}^2 |(- \im \grad) \psi|^2 \right] + \frac{b}{2} |\psi|^4 \right\}.
\label{Eq:SGL}
\end{equation}
Here $\psi (\rbm)$ [$= V^{-1/2} \sum_{\qbm} a_{\qbm 0} \exp (\im \qbm \cdot \rbm)$] is the classical order-parameter field in the coordinate representation, $\epsilon$ [$= (T - \Tc) / \Tc$] is the dimensionless temperature measured from the Gaussian critical temperature $\Tc$, and $\xiGL$ is the bare GL coherence length.
In the following, since only the classical fluctuation described by $a_{\qbm 0}$ is considered, we omit the suffix denoting the Matsubara index so that $a_{\qbm 0}$ is simply expressed as $a_{\qbm}$.
The coefficients in Eq.~(\ref{Eq:SGL}) are expressed as
\begin{eqnarray}
a &=& \frac{U}{4 \Tc V} \sum_{\kbm} {\varphi_k}^2 \left\{ Y_k - \frac{\del \mu}{\del T} \left[ \frac{\Tc}{\xi_k} Y_k - 2 \left( \frac{\Tc}{\xi_k} \right)^2 X_k \right] \right\}, \no \\
\label{Eq:GLCoeff_a} \\
b &=& \frac{U^2 \Tc}{V} \sum_{\kbm} {\varphi_k}^4 \left( \frac{X_k}{4 {\xi_k}^3} - \frac{Y_k}{8 \Tc {\xi_k}^2} \right),
\label{Eq:GLCoeff_b} \\
{\xiGL}^2 &=& \frac{U}{a V} \sum_{\kbm} {\varphi_k}^2 \frac{1}{16 m {\xi_k}^2} \left( X_k - \frac{\xi_k}{2 \Tc} Y_k + \frac{\xi_k k^2}{6 m {\Tc}^2} X_k Y_k \right), \no \\
\label{Eq:GLCoeff_xi}
\end{eqnarray}
where $X_k = \tanh [ \xi_k / (2 \Tc) ]$ and $Y_k = \text{sech}^2 [ \xi_k / (2 \Tc) ]$.
$\mu$ and $\del \mu / \del T$ appearing in Eqs.~(\ref{Eq:GLCoeff_a})-(\ref{Eq:GLCoeff_xi}) are assumed to take their values at $\Tc$.

\subsection{Zero-field SCF effect on specific heat}
\label{SubSec:Zero-field_SCF_effect_on_specific_heat}

To calculate the SCF-induced specific heat in zero magnetic field, we have to incorporate the mode-coupling effect in some manner.
To this end, we use the variational method~\cite{Adachi_Ikeda_2016} equivalent to the Hartree-Fock approximation, combined with the use of an effective high-energy cutoff of the SCF modes~\cite{Carballeira_Mosqueira_2000}.
In the following, we explain the details of this treatment.

We divide the GL action ($\SGL$) into two parts as
\begin{eqnarray}
\SGL &=& \SGL^0 + \SGL^1 \no \\
&=& \int \drm^3 \rbm \, a\left[ \eta |\psi|^2 + {\xiGL}^2 |(-\im \grad) \psi|^2 \right] \no \\
&& + \int \drm^3 \rbm \, \left[ a (\epsilon - \eta) |\psi|^2 + \frac{b}{2} |\psi|^4 \right],
\end{eqnarray}
where $\eta$ is a variational parameter corresponding to the renormalized mass of the SCF.
By tuning $\eta$ to optimize a trial free-energy density given below, we expect that $\SGL^0$ will be dominant while $\SGL^1$ will be a small perturbation~\cite{Jiang_Li_2014}.
The free-energy density $f(\epsilon)$ can be estimated as
\begin{equation}
f = f_0 - \frac{\Tc}{V} \ln \langle \e^{-\SGL^1} \rangle_\text{GL}^0 \leq f_0 + \frac{\Tc}{V} \langle \SGL^1 \rangle_\text{GL}^0 \equiv \ftri,
\label{Eq:VariationalMethod}
\end{equation}
where $\ftri (\epsilon; \eta)$ is a trial free-energy density, which should be optimized (\textit{i.e.}, minimized) with respect to $\eta$.
Here $f_0 (\eta)$ is the contribution from $\SGL^0$, and $\langle \cdots \rangle_\text{GL}^0$ is the average with respect to $\SGL^0$.

As mentioned above, the GL action $\SGL$ is meaningful only for the low-energy SCF.
Since $\SGL$ is isotropic in real space, it is natural that we should impose an isotropic cutoff in the momentum space.
Accordingly, we simply introduce a high-energy (and short-wavelength) cutoff ${\xi_0}^2 {\qc}^2$.
In other words, we restrict the SCF modes to those satisfying
\begin{equation}
{\xi_0}^2 q^2 \leq {\xi_0}^2 {\qc}^2 \equiv c^2,
\label{Eq:CutoffCondition}
\end{equation}
where $c = \mathcal{O} (1)$.
Using this cutoff, we can obtain the explicit form of $f_0 (\eta)$ as follows:
\begin{eqnarray}
f_0 (\eta) &=& - \frac{\Tc}{V} \ln \int \left( \prod_{\qbm} \frac{\drm \text{Re}a_{\qbm} \, \drm \text{Im}a_{\qbm}}{\pi} \right) \e^{-\SGL^0} \no \\
&=& \frac{\Tc}{2 \pi^2 {\xiGL}^3} \left\{ - \frac{2}{9} c^3 + \frac{2}{3} c \eta - \frac{2}{3} \eta^{3/2} \arctan \frac{c}{\sqrt{\eta}} \right. \no \\
&& \left. + \frac{1}{3} c^3 \ln \left[ a \left( c^2 + \eta \right) \right] \right\}.
\label{Eq:f0}
\end{eqnarray}
The optimizing equation $\del \ftri / \del \eta = 0$ can be transformed into
\begin{equation}
\eta = \epsilon + \frac{2 b}{a^2 \Tc} \frac{\del f_0}{\del \eta}.
\label{Eq:OptimizingEq}
\end{equation}
Combining Eq.~(\ref{Eq:f0}) and (\ref{Eq:OptimizingEq}), we explicitly obtain the optimizing equation:
\begin{equation}
\eta = \epsilon + \frac{2}{\pi} \sqrt{Gi} \left( c - \sqrt{\eta} \arctan \frac{c}{\sqrt{\eta}} \right),
\label{Eq:OptimizingEq2}
\end{equation}
The Ginzburg number $Gi$ in Eq.~(\ref{Eq:OptimizingEq2}) represents the mode-coupling strength and is defined as
\begin{equation}
Gi = \left( \frac{b}{2 \pi a^2 {\xiGL}^3} \right)^2 = \left( \frac{1}{2 \pi \Delta c_V {\xiGL}^3} \right)^2,
\label{Eq:GinzburgNum}
\end{equation}
where $\Delta c_V$ is the mean-field specific-heat jump at $\Tc$.

The trial free-energy density $\ftri (\epsilon, \eta)$ can be rewritten as
\begin{equation}
\ftri (\epsilon, \eta) = f_0 + (\epsilon - \eta) \frac{\del f_0}{\del \eta} + \frac{b}{\Tc a^2} \left( \frac{\del f_0}{\del \eta} \right)^2.
\label{Eq:ftri}
\end{equation}
The optimized free-energy density $\fopt (\epsilon)$ can be obtained from the combination of Eqs.~(\ref{Eq:OptimizingEq}) and (\ref{Eq:ftri}).
The entropy density $s (\epsilon)$ and specific heat $c_V (\epsilon)$ are respectively given as
\begin{equation}
s = - \frac{\del \fopt}{\del T} = - \frac{1}{2 \pi^2 {\xiGL}^3} \left( c - \sqrt{\eta^*} \arctan \frac{c}{\sqrt{\eta^*}} \right)
\end{equation}
and $c_V = \Tc \del s / \del T$,
where $\eta^*$ is the solution of the optimizing equation [Eq.~(\ref{Eq:OptimizingEq2})].
We note that small temperature dependence irrelevant to $\epsilon$ is neglected.

\subsection{Renormalization of critical temperature and coherence length at zero temperature}
\label{SubSec:Renormalization_of_the_critical_temperature_and_coherence_length}

Now we examine how the critical temperature and the coherence length are renormalized via the mode coupling between the SCF.
The mode coupling decreases the critical temperature $\Tc$, which is defined in the Gaussian approximation, down to a renormalized critical temperature $\TcR$.
In our formalism, $\TcR$ is determined from Eq.~(\ref{Eq:OptimizingEq2}) by setting the renormalized mass $\eta$ to zero:
\begin{equation}
\TcR = \left( 1 - \frac{2 c}{\pi} \sqrt{Gi} \right) \Tc.
\end{equation}
When $T$ is so close to $\TcR$ that the condition $\eta \ll c$ is satisfied, Eq.~(\ref{Eq:OptimizingEq2}) can be rewritten as
\begin{equation}
\eta = \frac{T - \TcR}{\Tc} - \sqrt{Gi} \sqrt{\eta}.
\end{equation}
Therefore, the renormalized mass $\eta$ is asymptotically given as
\begin{equation}
\eta \simeq
\left\{
\begin{array}{l}
\displaystyle \frac{T - \TcR}{\Tc} \ \ \left( Gi \ll \frac{T - \TcR}{\Tc} \ll c \right) \vspace{5pt} \\
\displaystyle \frac{1}{Gi} \left( \frac{T - \TcR}{\Tc} \right)^2 \ \ \left( \frac{T - \TcR}{\Tc} \ll Gi \right).
\end{array}
\right.
\label{Eq:eta_case}
\end{equation}
Thus when $Gi \ll (T - \TcR) / \Tc \ll c$, the dominant part of the GL action $\SGL^0$ can be approximated as follows:
\begin{eqnarray}
\SGL^0 &\simeq& \int \drm^3 \rbm \left( \frac{\TcR}{\Tc} \right) a \left[ \frac{T - \TcR}{\TcR} |\psi|^2 + \frac{\Tc}{\TcR} {\xiGL}^2 |(- \im \grad) \psi|^2 \right] \no \\
&=& \int \drm^3 \rbm \, \aR \left[ \epsilonR |\psi|^2 + {\xiR}^2 |(- \im \grad) \psi|^2 \right],
\end{eqnarray}
where the renormalized parameters are given as
\begin{eqnarray}
\aR = \left( \frac{\TcR}{\Tc} \right) a, \\
\epsilonR = \frac{T - \TcR}{\TcR}, \\
{\xiR}^2 = \frac{\Tc}{\TcR} {\xiGL}^2.
\label{Eq:xiR}
\end{eqnarray}
From Eq.~(\ref{Eq:xiR}), we find that the coherence length is renormalized along with the renormalization  of the critical temperature.
Thus, it is not the bare depairing field (the so-called upper critical field) $\Bctwo (T)$ [$= \phi_0 (\Tc - T) / (2 \pi {\xi_0}^2 \Tc)$] but the corresponding renormalized one
\begin{equation}
B_\text{c2R} (T) \equiv \frac{\phi_0 (\TcR - T)}{2 \pi {\xiR}^2 \TcR}
\label{Eq:Bc2R}
\end{equation}
that is estimated from conventional experiments, where $\phi_0$ is the flux quantum.
That is, since the coherence length determined experimentally will be not $\xiGL$ but $\xiR$, one should pay attention to the difference between $\xiGL$ and $\xiR$ especially when the mode coupling is important as in the BCS-BEC-crossover regime.
Similar renormalization of the coherence length due to the reduction of the critical temperature has also been stressed in the context of underdoped cuprates~\cite{Ikeda_2002}.

According to Eq.~(\ref{Eq:eta_case}), the temperature dependence of the correlation length $\xi_\text{R}(T)=\xiR/\sqrt{\eta (T)}$ defined above the superconducting transition is changed on approaching $\TcR$, and we have $\xi_\text{R} (T) \sim (T - \TcR)^{-1}$ in the vicinity of $\TcR$.
Consequently, within the present Hartree-Fock approach, the critical behaviors of physical quantities may be remarkably different from those in the Gaussian approximation where $\xi_\text{R} (T) \sim (T - \TcR)^{-1/2}$.
As an illustration, we focus here on the diamagnetic susceptibility $\chidia$ and the specific heat $c_V$.
Noting that the singular part of the free-energy density behaves like $- \Tc / [\xi_\text{R}(T)]^3$ and that a change of the flux density carries the factor $[\xi_\text{R}(T)]^2$ due to the gauge invariance while $c_V$ is the second derivative of the free-energy density with respect to $T$, one finds that $\chidia \propto (T - \TcR)^{-1}$ while $c_V$ saturates a finite value in the vicinity of $\TcR$. 

\subsection{SCF effect on specific heat and diamagnetism in magnetic fields}
\label{SubSec:SCF_effect_on_specific_heat_and_diamagnetism_in_magnetic_fields}

In non-zero magnetic fields, $- \im \grad$ in the GL action [Eq.~(\ref{Eq:SGL})] needs to be replaced with $- \im \grad + 2 \pi \Abm / \phi_0$.
Here, $\Abm (\rbm)$ $[= (0, B x, 0)]$ is the vector potential in the Landau gauge.
As far as the paramagnetic pair-breaking effect is negligible, the resulting GL action will correctly describe the low-energy SCF around $\Tc$ in magnetic fields.

To estimate the SCF-induced diamagnetism and specific heat in magnetic fields, we use the variational method combined with the introduction of a certain cutoff as in the zero-field case, which is explained in the following.

We first divide the action $\SGL$ in magnetic fields into two parts:
\begin{eqnarray}
\SGL &=& \SGL^0 + \SGL^1 \no \\
&=& \int \drm^3 \rbm \, a \left[ \eta |\psi|^2 + {\xi_0}^2 \left| \left( - \im \grad + \frac{2 \pi}{\phi_0} \Abm \right) \psi \right|^2 \right] \no \\
&& + \int \drm^3 \rbm \left[ a \left( \epsilon - \eta \right) |\psi|^2 + \frac{b}{2} |\psi|^4 \right].
\end{eqnarray}
Then, in quite the same way as in the zero-field case, the trial free-energy density $\ftri$ is estimated with $f_0$.
We stress that the equations determining the optimized free-energy density $\fopt$ [Eqs.~(\ref{Eq:VariationalMethod}), (\ref{Eq:OptimizingEq}), and (\ref{Eq:ftri})] are just the same as in the zero-field case (the explicit form of $f_0$ is explained in the next paragraph).
However, the order-parameter field in magnetic fields needs to be expanded as
\begin{equation}
\psi (\rbm) = \sum_{N, q_y, q_z} b_{N q_y q_z} f_{N q_y} (x) \frac{\e^{\im q_y y}}{\sqrt{L_y}} \frac{\e^{\im q_z z}}{\sqrt{L_z}},
\label{Eq:OP_Mag}
\end{equation}
where $N$ is the Landau-level index, $f_{N q_y} (x)$ is the $N$th-Landau-level eigen function, and $b_{N q_y q_z}$ is the expansion coefficient describing the SCF mode in magnetic fields.
By using this representation, $\SGL^0$ is written as
\begin{equation}
\SGL^0 = \sum_{N, q_y, q_z} a \left[ \eta + 2 h \left( N + \frac{1}{2} \right) + {\xi_0}^2 {q_z}^2 \right] |b_{N q_y q_z}|^2,
\label{Eq:S0_Mag}
\end{equation}
where $h$ [$= 2 \pi {\xi_0}^2 B / \phi_0 \equiv B / \Bctwo (0)$] is a dimensionless magnetic field.
As integral variables for $f_0$, we can choose $\{ b_{N q_y q_z}, b^*_{N q_y q_z} \}$ instead of $\{ a_{\qbm}, a^*_{\qbm} \}$ [\textit{cf}. Eq.~(\ref{Eq:f0})]:
\begin{equation}
f_0 = - \frac{\Tc}{V} \ln \int \left( \prod_{N, q_y, q_z} \frac{\drm \text{Re}b_{N q_y q_z} \, \drm \text{Im}b_{N q_y q_z}}{\pi} \right) \e^{-\SGL^0}.
\label{Eq:f0_Mag}
\end{equation}

Since $\SGL$ cannot correctly describe the high-energy SCF, we should introduce a high-energy cutoff ${\xi_0}^2 {\qc}^2$.
Here we restrict the SCF modes to those satisfying
\begin{equation}
\left\{
\begin{array}{l}
2 h (N + 1) + {\xi_0}^2 {q_z}^2 \leq {\xi_0}^2 {\qc}^2 \equiv c^2 \vspace{5pt} \\
{\xi_0}^2 {q_z}^2 \leq {\xi_0}^2 {\qc}^2.
\end{array}
\right.
\label{Eq:CutoffCondition_Mag}
\end{equation}
We note that the factor $N + 1$ is different from $N + 1 / 2$ appearing in Eq. (\ref{Eq:S0_Mag}).
Combining Eqs.~(\ref{Eq:S0_Mag}), (\ref{Eq:f0_Mag}), and (\ref{Eq:CutoffCondition_Mag}) leads to the following explicit form of $f_0$:
\begin{equation}
f_0 (h; \eta) = \frac{\Tc}{2 \pi^2 {\xi_0}^3} \left[ \frac{c^3}{3} \ln (2 a h) + h I_0 (h; \eta) \right],
\label{Eq:f0_Mag_Explicit}
\end{equation}
where
\begin{equation}
I_0 (h; \eta) = \int_0^c \drm x \left[ \ln \Gamma \left( \frac{\eta + c^2}{2 h} + \frac{1}{2} \right) - \ln \Gamma \left( \frac{\eta + x^2}{2 h} + \frac{1}{2} \right) \right].
\end{equation}
Here $\Gamma (x)$ is the gamma function.
With the use of Eqs.~(\ref{Eq:OptimizingEq}) and (\ref{Eq:f0_Mag_Explicit}), the explicit form of the optimizing equation is given by
\begin{equation}
\eta = \epsilon + \frac{1}{\pi} \sqrt{Gi} I_1 (h; \eta),
\label{Eq:OptimizingEq_Mag_Explicit}
\end{equation}
where
\begin{equation}
I_1 (h; \eta) = \int_0^c \drm x \left[ \psi \left( \frac{\eta + c^2}{2 h} + \frac{1}{2} \right) - \psi \left( \frac{\eta + x^2}{2 h} + \frac{1}{2} \right) \right].
\label{Eq:I1}
\end{equation}
Here $\psi (x) = \drm \ln \Gamma (x) / \drm x$ is the digamma function.

\begin{figure*}[htbp]
\begin{tabular}{c}
\begin{minipage}[t]{0.47\hsize}
\centering
\includegraphics[scale = 0.56, clip]{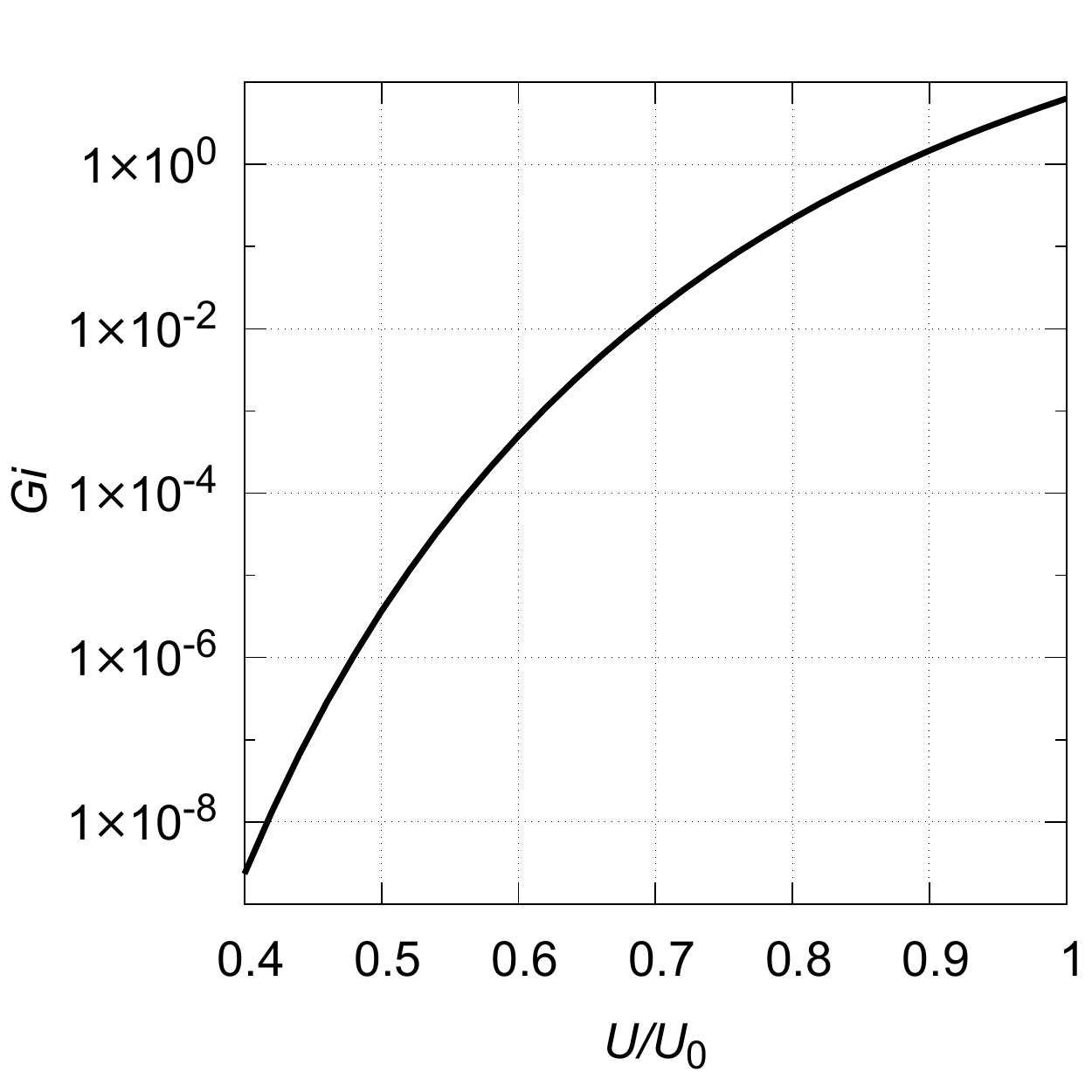}
\caption{
The Ginzburg number $Gi$ as a function of the interaction strength $U$.
The vertical scale is logarithmic.
}
\label{Fig:Gi}
\end{minipage}
\begin{minipage}[t]{0.03\hsize}
\hspace{0pt}
\end{minipage}
\begin{minipage}[t]{0.47\hsize}
\centering
\includegraphics[scale = 0.56, clip]{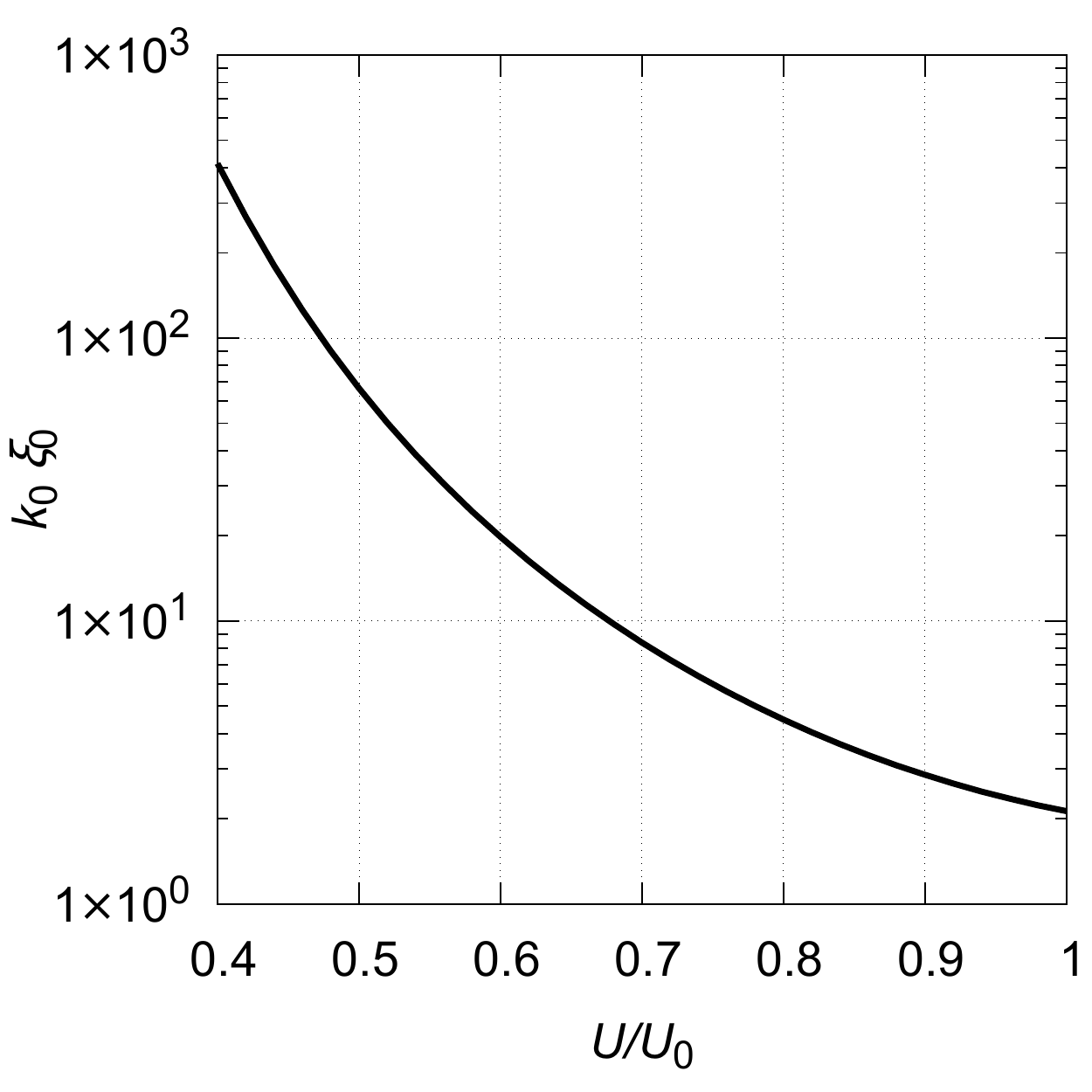}
\caption{
The bare coherence length $\xi_0$ as a function of the interaction strength $U$.
The vertical scale is logarithmic.
}
\label{Fig:xi0}
\end{minipage}
\end{tabular}
\end{figure*}

Now we explain why we choose the above-mentioned cutoff condition [Eq.~(\ref{Eq:CutoffCondition_Mag})].
Considering the zero-field limit ($h \rightarrow 0$) in Eq.~(\ref{Eq:f0_Mag_Explicit}) and using the asymptotic form of the Gamma function, we obtain the right-hand side of Eq.~(\ref{Eq:f0}).
As mentioned above, moreover, the equations determining the optimized free-energy density $\fopt$ are just the same as in the zero-field case and given by Eqs.~(\ref{Eq:VariationalMethod}), (\ref{Eq:OptimizingEq}), and (\ref{Eq:ftri}).
In other words, \textit{we can consistently reproduce the zero-field isotropic behavior if we adopt the cutoff condition given by Eq.~(\ref{Eq:CutoffCondition_Mag})}.
This is the reason why we choose Eq.~(\ref{Eq:CutoffCondition_Mag}).

The entropy density $s (\epsilon, h)$ is calculated as
\begin{equation}
s = - \frac{\del \fopt}{\del T} = - \frac{1}{4 \pi^2 {\xi_0}^3} I_1 (h; \eta^*),
\end{equation}
where $\eta^*$ is the solution of the optimizing equation [Eq.~(\ref{Eq:OptimizingEq_Mag_Explicit})].
The specific heat $c_V (\epsilon, h)$ is calculated as $c_V = \Tc \del s / \del T$.

The magnetization $\Mdia (\epsilon, h)$ is estimated as
\begin{equation}
\Mdia = - \frac{\del \fopt}{\del B} = - \frac{\Tc}{\pi \phi_0 \xi_0} \left[ \frac{c^3}{3 h} + I_0 (h; \eta^*) - I_3 (h; \eta^*) \right],
\end{equation}
where
\begin{eqnarray}
&& I_3 (h; \eta) \no \\
&& = \int_0^c \drm x \left[ \frac{\eta + c^2}{2 h} \psi \left( \frac{\eta + c^2}{2 h} + \frac{1}{2} \right) - \frac{\eta + x^2}{2 h} \psi \left( \frac{\eta + x^2}{2 h} + \frac{1}{2} \right) \right]. \no \\
\end{eqnarray}
The diamagnetic susceptibility $\chidia (\epsilon, h)$ is simply defined as $\chidia = \mu_0 \Mdia / B$, where $\mu_0$ is the vacuum permeability.

Before finishing this section, we note some comments.
If we consider the zero-field case ($h \rightarrow 0$) in addition to neglecting both of the mode-coupling effect ($\eta \rightarrow \epsilon$) and the cutoff effect ($c \rightarrow \infty$), $c_V$ approaches the familiar result in the Gaussian approximation:
\begin{equation}
c_V \rightarrow \frac{1}{8 \pi {\xi_0}^3} \frac{1}{\sqrt{\epsilon}}.
\label{Eq:SpecificHeat_Gaussian}
\end{equation}
On the other hand, if we neglect both of the mode-coupling effect ($\eta \rightarrow \epsilon$) and the cutoff effect ($c \rightarrow \infty$) while keeping the magnetic field finite, we obtain
\begin{equation}
\Mdia \rightarrow - \frac{\Tc}{\pi \phi_0 \xi_0} \int_0^\infty \drm x \Upsilon \left( \frac{\epsilon + x^2}{2 h} + \frac{1}{2} \right),
\end{equation}
where
\begin{equation}
\Upsilon (x) = - \ln \Gamma (x) + \left( x - \frac{1}{2} \right) \psi (x) - x + \frac{1}{2} \left[ 1 + \ln (2 \pi) \right].
\end{equation}
This equation is equivalent to the well-known Prange's result~\cite{Prange_1970}.

\begin{figure*}[htbp]
\begin{tabular}{c}
\begin{minipage}[t]{0.47\hsize}
\centering
\includegraphics[scale = 0.56, clip]{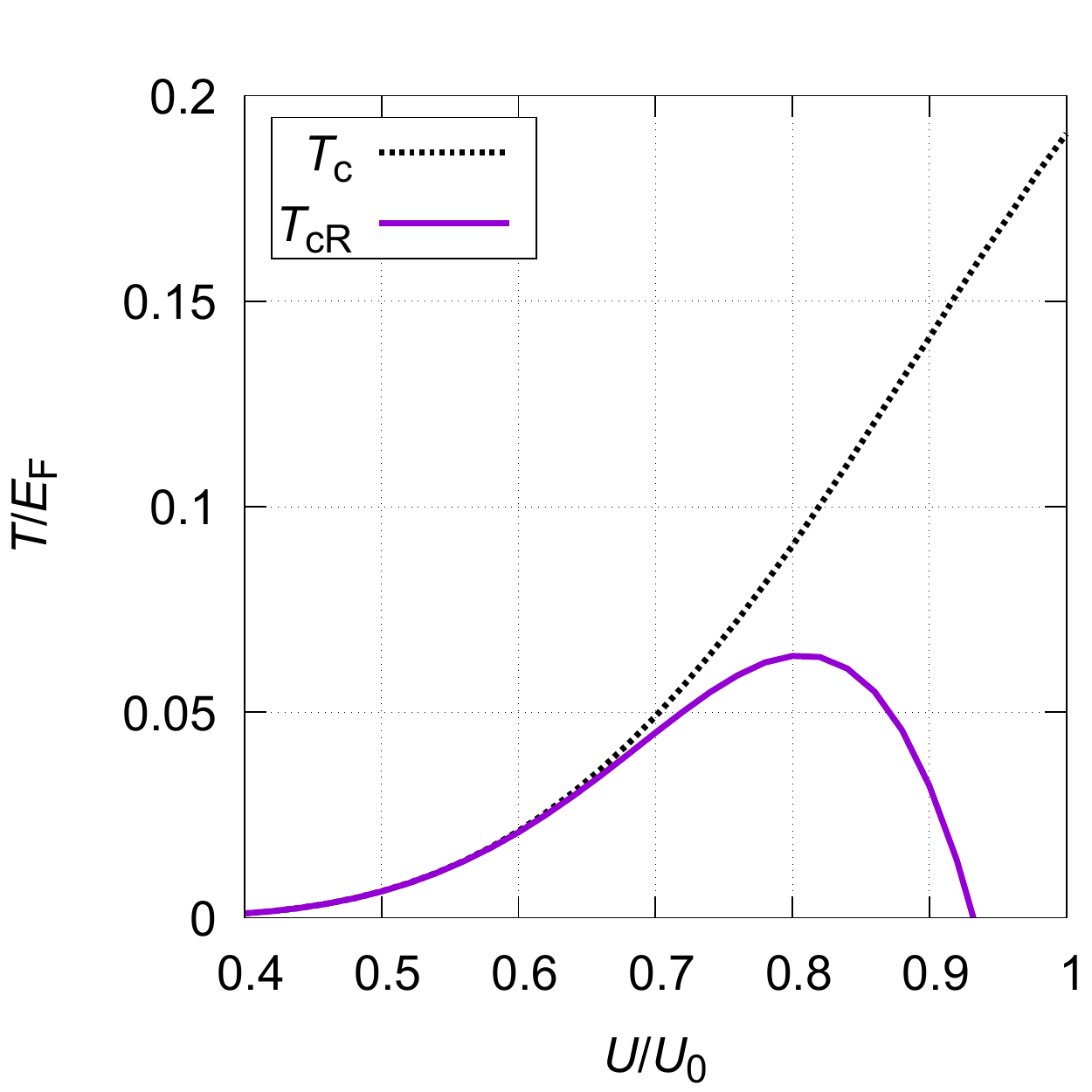}
\caption{
The bare critical temperature $\Tc$ (black dotted line) and the renormalized critical temperature $\TcR$ (purple solid line) as a function of the interaction strength $U$.
Our theoretical approach seems to be valid when $U / U_0 \leq 0.8$.
}
\label{Fig:Tc}
\end{minipage}
\begin{minipage}[t]{0.03\hsize}
\hspace{0pt}
\end{minipage}
\begin{minipage}[t]{0.47\hsize}
\centering
\includegraphics[scale = 0.56, clip]{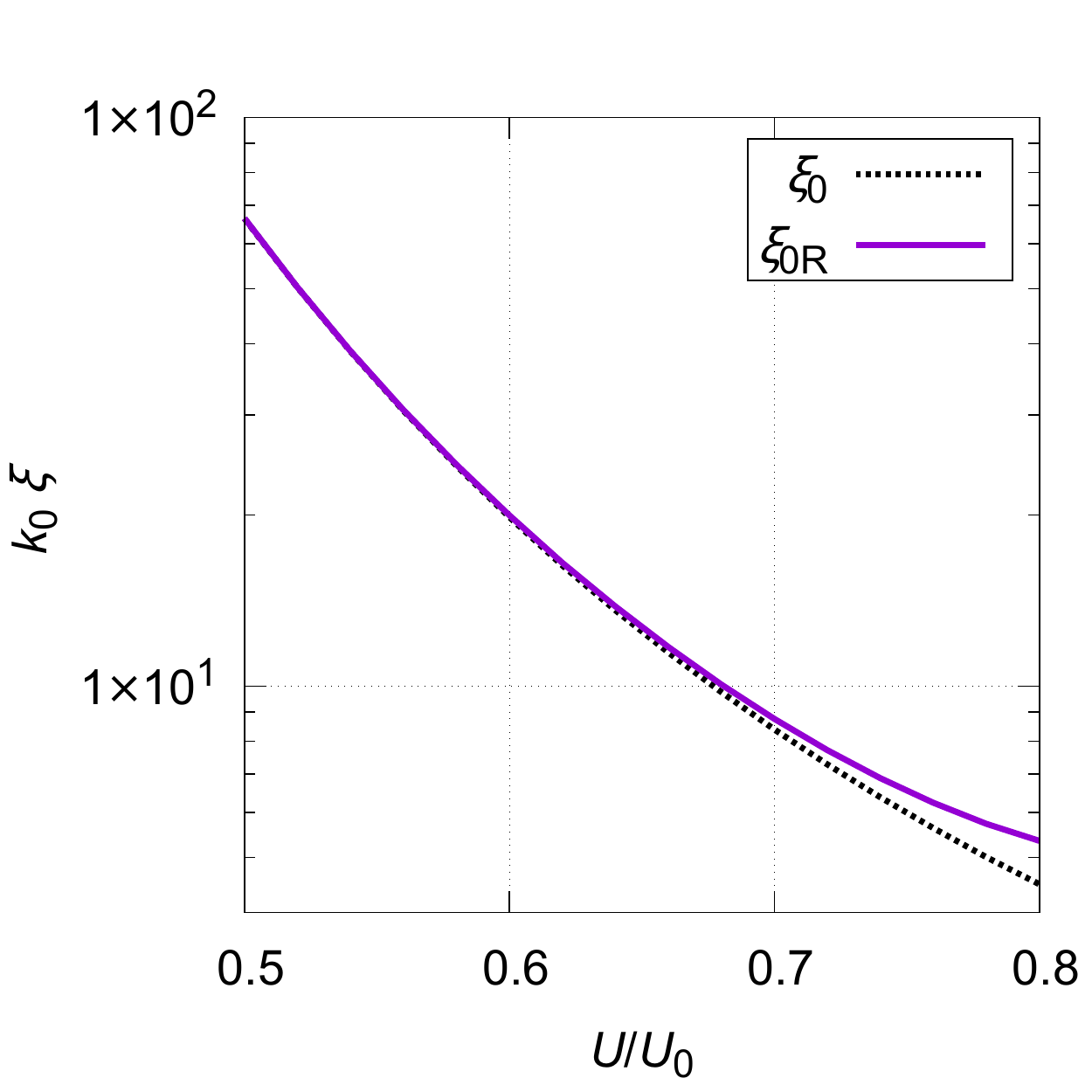}
\caption{
The bare coherence length $\xi_0$ (black dotted line) and the renormalized coherence length $\xi_\text{0R}$ (purple solid line) as a function of the interaction strength $U$.
The vertical scale is logarithmic.
}
\label{Fig:xi0R}
\end{minipage}
\end{tabular}
\end{figure*}

\section{Results}
\label{Sec:Results}

In the following, we explain the obtained results.
First, we fix some parameters to investigate the BCS-BEC-crossover regime as well as the BCS (or weak-coupling) regime.
Second, the renormalization of both the critical temperature and the coherence length is shown to be quantitatively large in the BCS-BEC-crossover regime.
Then, we demonstrate that the obtained specific heat and diamagnetic susceptibility can exceed in magnitude their Gaussian-approximation values.
Next, we explain that the so-called lowest-Landau-level scaling, observed in high-$\Tc$ cuprates' data in tesla range, can break down in the BCS-BEC-crossover regime due to an increase~\cite{Ikeda_1995} of $Gi$, which measures the strength of the mode coupling.
Finally, we illustrate that the crossing behavior of magnetization curves, often observed in 2D-like high-$\Tc$ cuprates, can still occur in the present 3D systems over a broad field range in the BCS-BEC-crossover regime, even though the lowest-Landau-level scaling is ill-defined.

\subsection{Parameters}
\label{SubSec:Parameters}

As a preliminary, we note the values of some parameters used in our numerical calculations.
First, by using the momentum cutoff [$k_0$ appearing in Eq.~(\ref{Eq:FormFactor})], we fix the total number density as $\ntot {k_0}^{-3} \simeq 0.007$, which corresponds to a relatively dilute electron system.
Next, the effective high-energy cutoff is fixed as $c = \xi_0 \qc = 1$, where $\xi_0$ is the bare coherence length and $\qc$ is the short-wavelength cutoff of the SCF mode.
There is a tendency for the obtained results to be qualitatively insensitive to $c$ as long as $c$ is small.

As mentioned in Sec.~\ref{Sec:Introduction}, the critical region or the Ginzburg number $Gi$ should be large in the BCS-BEC-crossover regime.
Thus to fix a typical value of attractive interaction $U$ in each of the BCS and BCS-BEC-crossover regime, we check the $U$ dependence of $Gi$~\footnote{We neglect the term proportional to $\del \mu / \del T$ in Eq.~(\ref{Eq:GLCoeff_a}) since this term is small compared to the other term at least in the interaction range $U / U_0 \lesssim 1$.} and obtain Fig.~\ref{Fig:Gi}, where $U$ is measured in units of $U_0$, the threshold value necessary to form a two-particle bound state (see Sec.~\ref{SubSec:Model}).
Figure~\ref{Fig:Gi} shows that $Gi$ rapidly increases as $U$ becomes large.
Reflecting the well-known expression of $\Tc$ valid in the BCS regime, $\Tc \sim \exp [- 1 / (N_\text{F} U)]$, where $N_\text{F}$ is the density of states at the Fermi surface, $\xi_0$ ($\propto {\Tc}^{-1}$) sharply decreases as $U$ increases at least in the BCS regime (Fig.~\ref{Fig:xi0}).
Since $Gi \propto {\xi_0}^{-6}$ in addition, the rapid increase of $Gi$ is mainly due to the sharp decrease of $\xi_0$.
From Fig.~\ref{Fig:Gi}, we assume that $U / U_0 = 0.5$ ($Gi \sim 10^{-6}$) and $U / U_0 = 0.8$ ($Gi \sim 0.2$) correspond to the values for typical BCS and BCS-BEC-crossover regimes, respectively.

\subsection{Critical temperature and coherence length}
\label{SubSec:Critical_temperature_and_coherence_length}

The bare critical temperature $\Tc$ and the critical temperature $\TcR$ renormalized by the mode coupling between SCF (see Sec.~\ref{SubSec:Renormalization_of_the_critical_temperature_and_coherence_length}) are shown in Fig.~\ref{Fig:Tc}.
Here the temperatures are measured in units of the Fermi energy $\EF$ [$= (3 \pi^2 \ntot)^{2 / 3} / (2 m)$].
We see from Fig.~\ref{Fig:Tc} that the renormalization (or lowering) of the critical temperature is enhanced in the BCS-BEC-crossover regime.
This is because of the strong mode coupling between SCF or the large value of $Gi$, as mentioned in Sec.~\ref{SubSec:Parameters}.
We also see that $\TcR$ starts to decrease as a function of $U$ when $U / U_0$ exceeds around $0.8$.
This may mean that our theoretical approach, the variational method on the GL action combined with the effective high-energy cutoff, is improper when $U / U_0 \gtrsim 0.8$.
In other words, we believe that our approach should be proper if $U / U_0 \lesssim 0.8$.
Therefore, we restrict our analysis of the SCF effects to the interaction range satisfying $U /U_0 \leq 0.8$.
We note that this interaction range includes the BCS ($U / U_0 = 0.5$) and BCS-BEC-crossover ($U / U_0 = 0.8$) cases defined in Sec.~\ref{SubSec:Parameters}.

Figure~\ref{Fig:xi0R} shows the renormalized coherence length $\xiR$ and the bare one $\xi_0$ within the range of the interaction satisfying $U / U_0 \leq 0.8$.
We see that the renormalized coherence length is elongated compared with the bare one especially in the BCS-BEC-crossover regime.

\begin{figure*}[htbp]
\begin{tabular}{c}
\begin{minipage}[t]{0.47\hsize}
\centering
\includegraphics[scale = 0.56, clip]{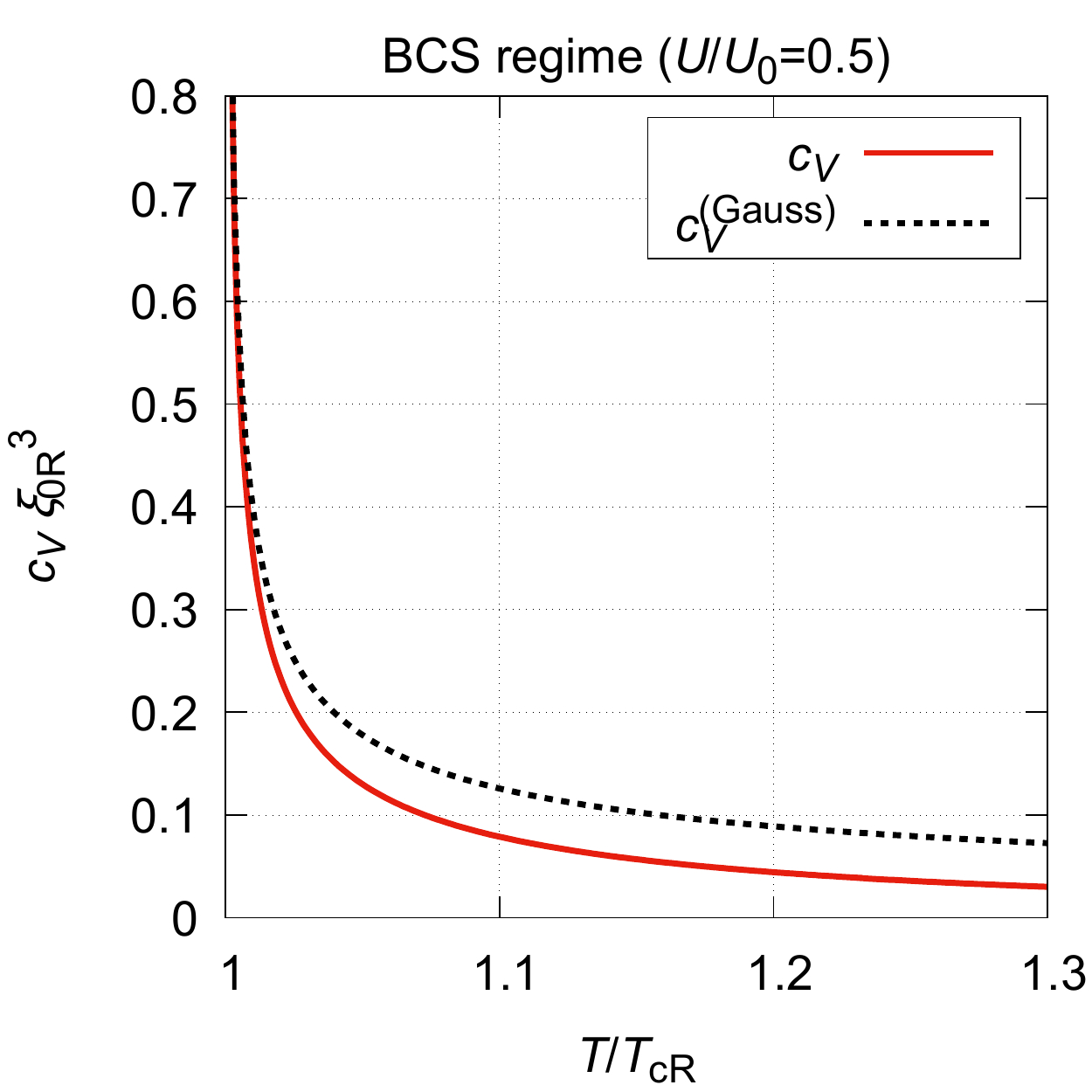}
\caption{
SCF-induced specific heat in the low-field limit as a function of temperature in the BCS regime.
The numerical result (red solid line) is shown with the result in the Gaussian approximation (black dotted line).
}
\label{Fig:cV_U05}
\end{minipage} 
\begin{minipage}[t]{0.03\hsize}
\hspace{0pt}
\end{minipage}
\begin{minipage}[t]{0.47\hsize}
\centering
\includegraphics[scale = 0.56, clip]{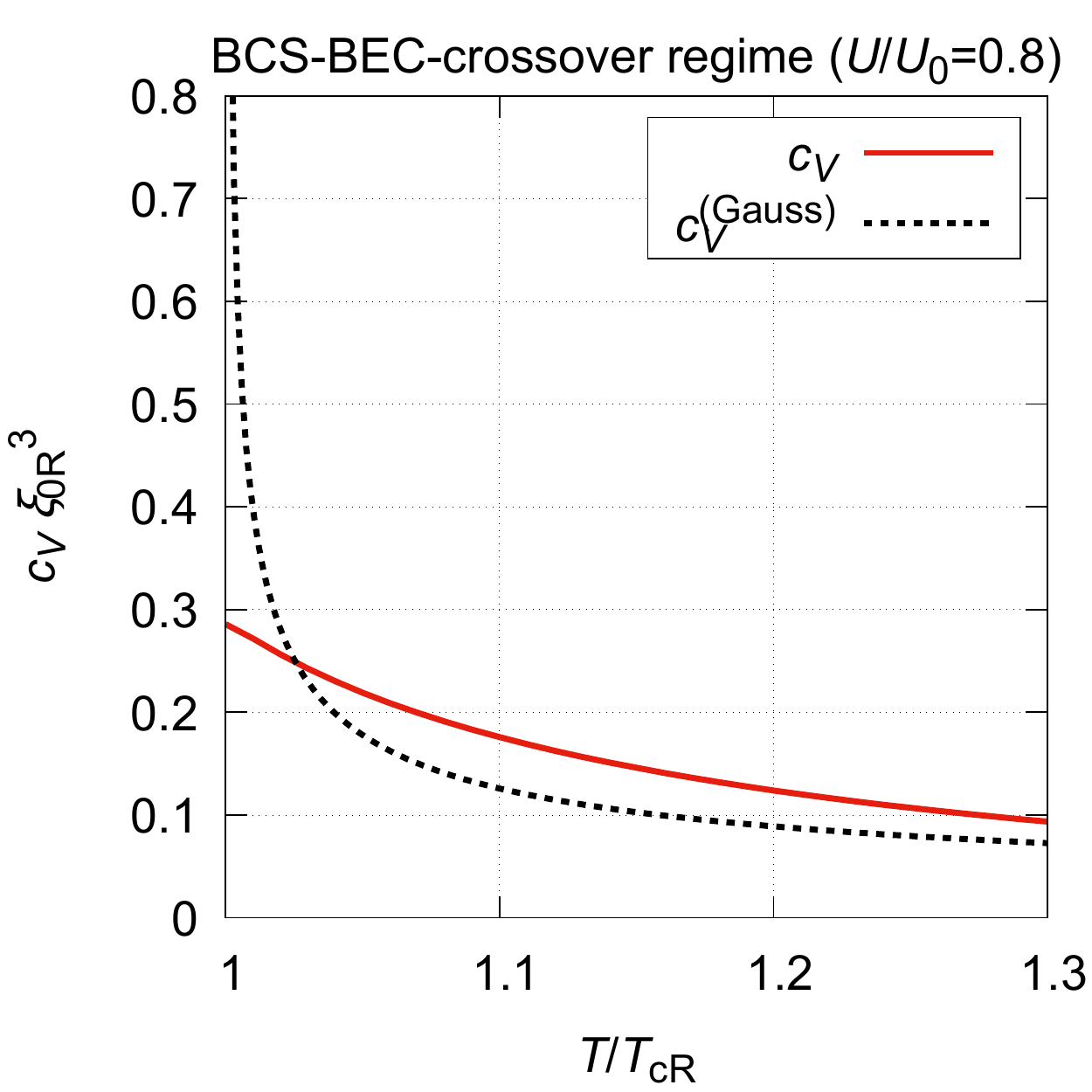}
\caption{
SCF-induced specific heat in the low-field limit as a function of temperature in the BCS-BEC-crossover regime.
The meaning of lines is the same as in Fig.~\ref{Fig:cV_U05}.
We note that, as explained at the end of subsection~\ref{SubSec:Renormalization_of_the_critical_temperature_and_coherence_length}, the renormalized specific heat saturates a finite value within the present method (see the red solid curve).
}
\label{Fig:cV_U08}
\end{minipage} \\
\begin{minipage}[t]{0.03\hsize}
\vspace{0pt}
\end{minipage} \\
\begin{minipage}[t]{0.47\hsize}
\centering
\includegraphics[scale = 0.56, clip]{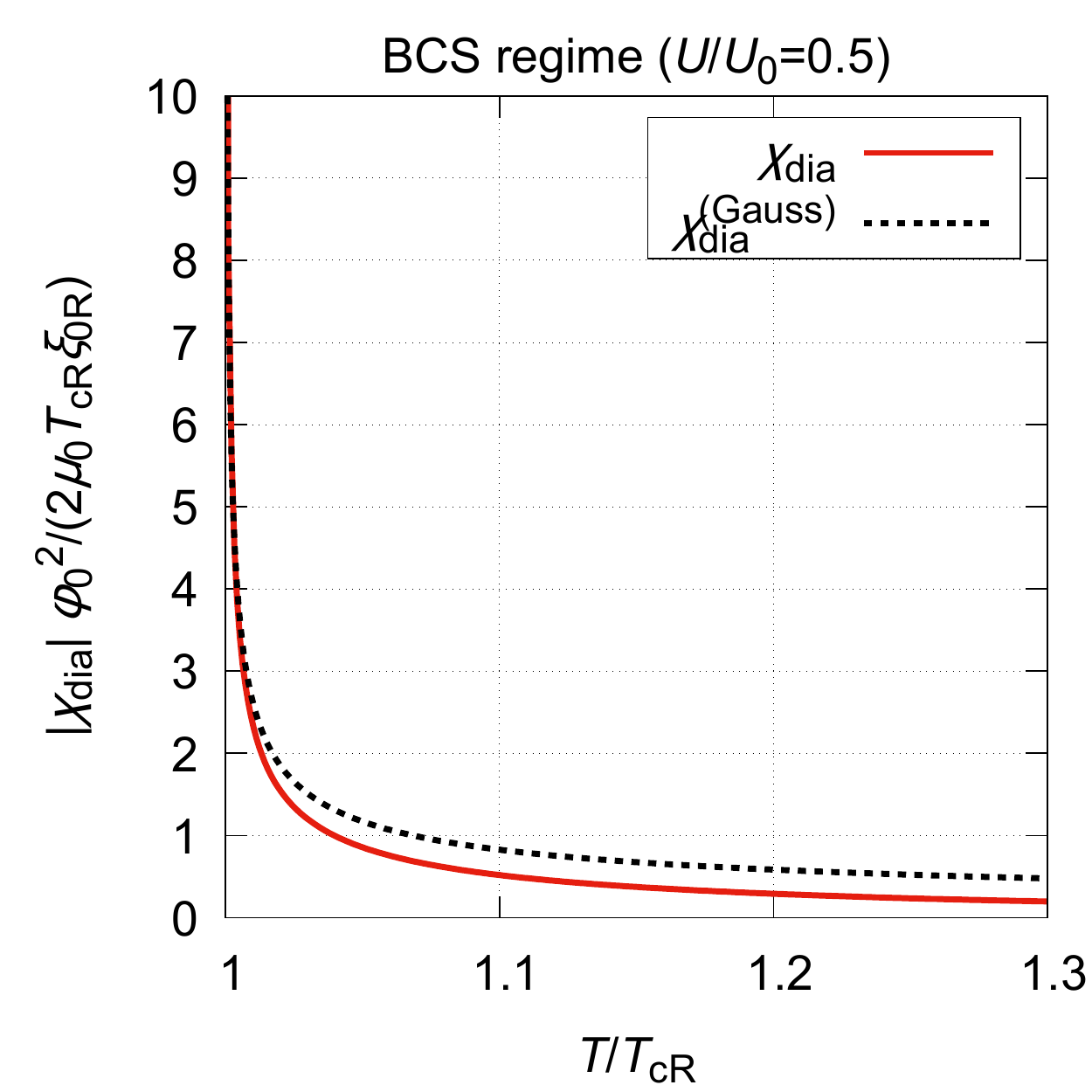}
\caption{
SCF-induced diamagnetic susceptibility in the low-field limit as a function of temperature in the BCS regime.
The numerical result (red solid line) is shown with the result in the Gaussian approximation (black dotted line).
We note that $\chidia$ is measured in units of $[{\phi_0}^2 / (2 \mu_0 \TcR \xiR)]^{-1}$.
}
\label{Fig:chi_U05}
\end{minipage}
\begin{minipage}[t]{0.03\hsize}
\hspace{0pt}
\end{minipage}
\begin{minipage}[t]{0.47\hsize}
\centering
\includegraphics[scale = 0.56, clip]{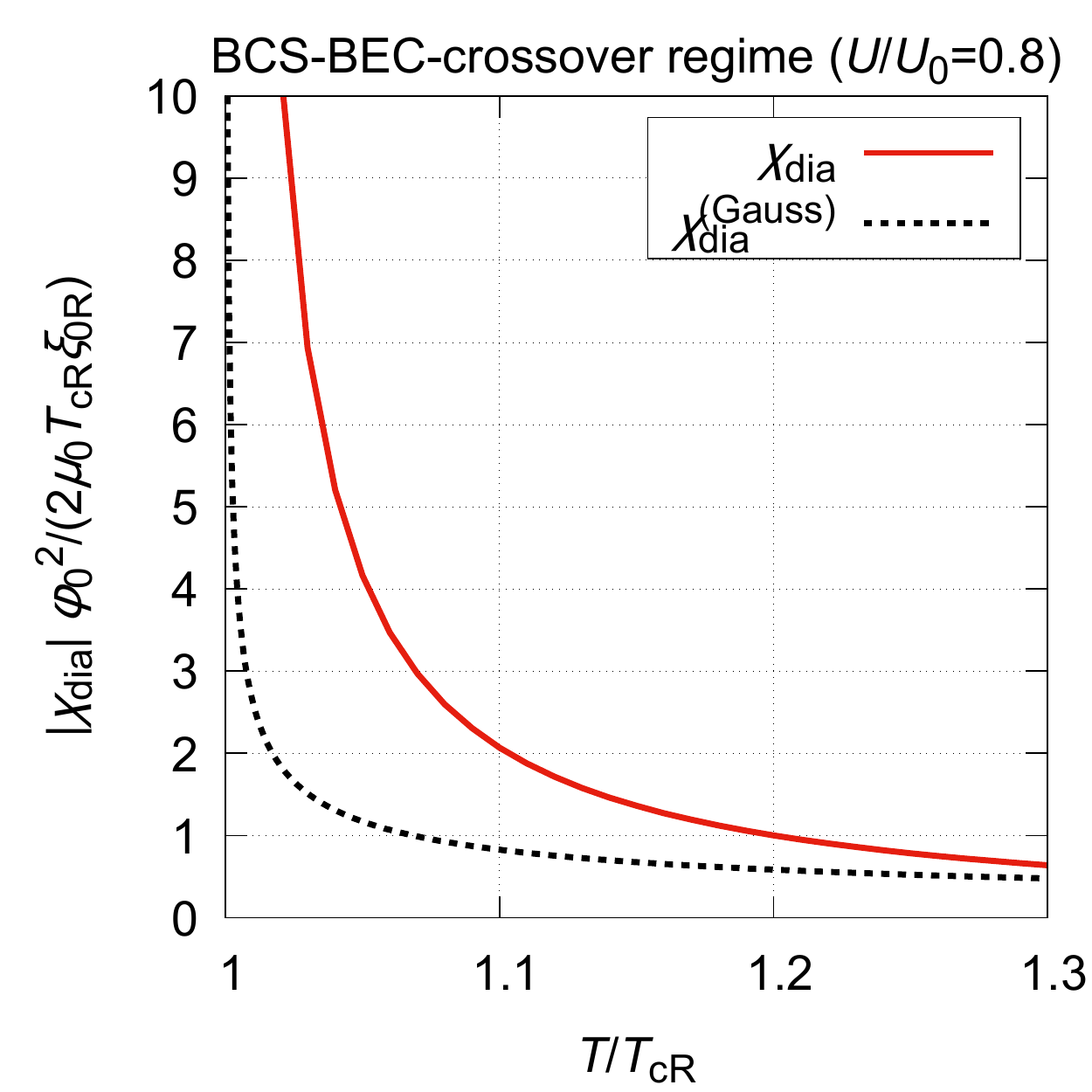}
\caption{
SCF-induced diamagnetic susceptibility in the low-field limit as a function of temperature in the BCS-BEC-crossover regime.
The meaning of lines is the same as in Fig.~\ref{Fig:chi_U05}.
}
\label{Fig:chi_U08}
\end{minipage}
\end{tabular}
\end{figure*}

\begin{figure*}[htbp]
\begin{tabular}{c}
\begin{minipage}[t]{0.47\hsize}
\centering
\includegraphics[scale = 0.56, clip]{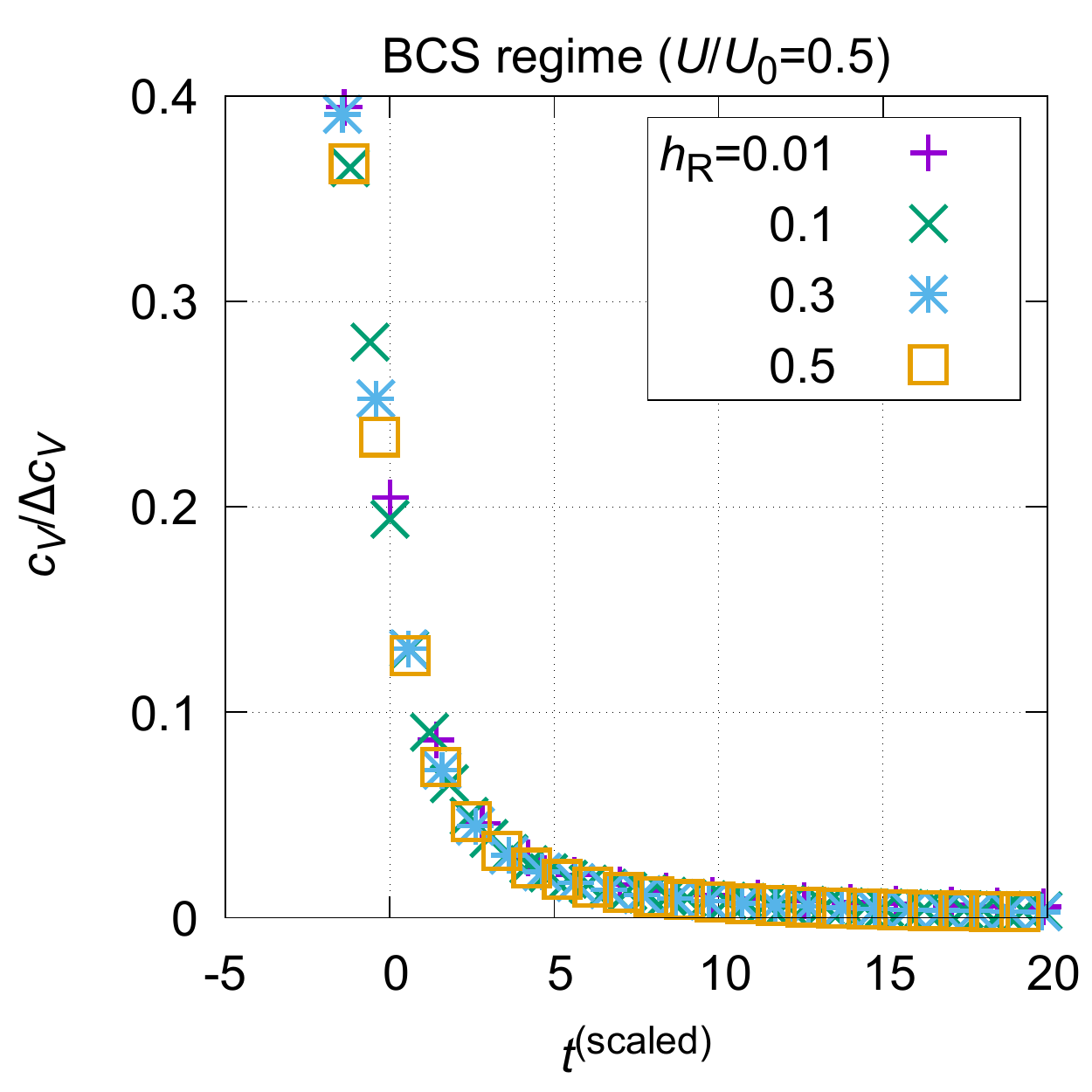}
\caption{
LLL-scaling plot of the SCF-induced specific heat in the BCS regime.
$t^\text{(scaled)}$ is the scaled temperature and $h_\text{R}$ is a magnetic field in units of $\BctwoR (0)$ [see Eq.~(\ref{Eq:Bc2R})].
}
\label{Fig:cV_Scaled_U05}
\end{minipage}
\begin{minipage}[t]{0.03\hsize}
\hspace{0pt}
\end{minipage}
\begin{minipage}[t]{0.47\hsize}
\centering
\includegraphics[scale = 0.56, clip]{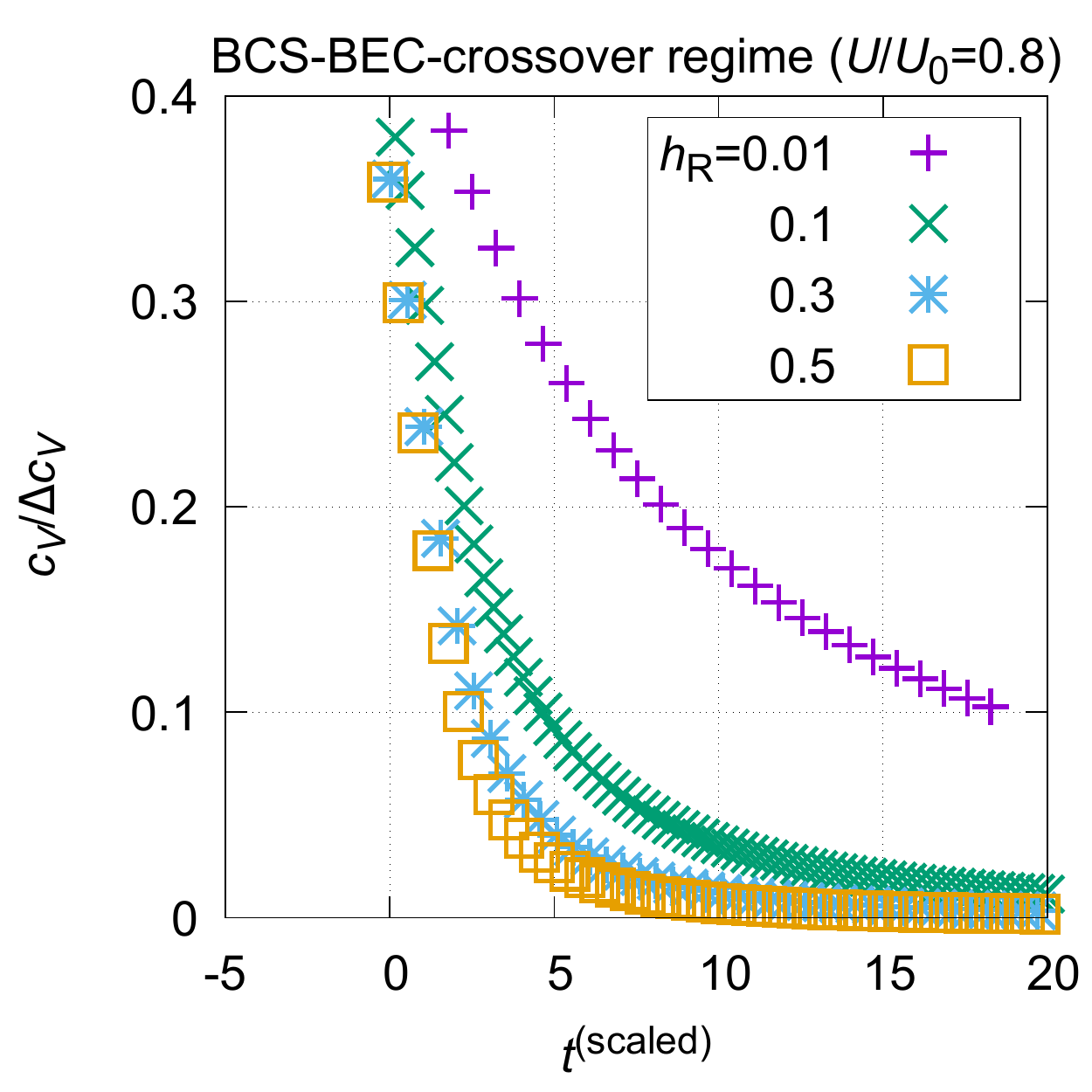}
\caption{
LLL-scaling plot of the SCF-induced specific heat in the BCS-BEC-crossover regime.
The symbols are used in the same way as in Fig.~\ref{Fig:cV_Scaled_U05}.
}
\label{Fig:cV_Scaled_U08}
\end{minipage} \\
\begin{minipage}[t]{0.03\hsize}
\vspace{0pt}
\end{minipage} \\
\begin{minipage}[t]{0.47\hsize}
\centering
\includegraphics[scale = 0.56, clip]{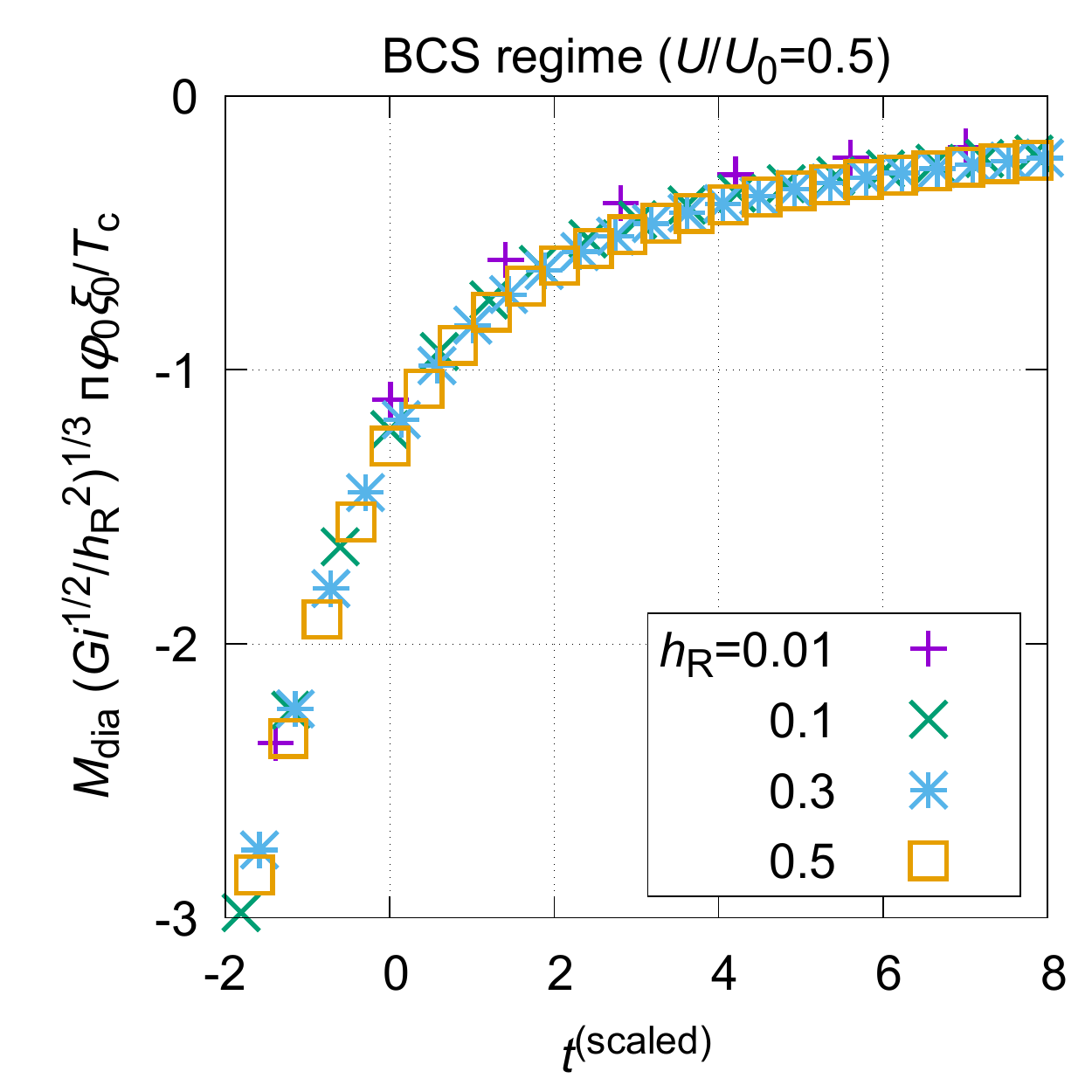}
\caption{
LLL-scaling plot of the SCF-induced magnetization in the BCS regime.
$t^\text{(scaled)}$ is the scaled temperature and $h_\text{R}$ is a magnetic field in units of $\BctwoR (0)$.
}
\label{Fig:Mdia_Scaled_U05}
\end{minipage}
\begin{minipage}[t]{0.03\hsize}
\hspace{0pt}
\end{minipage}
\begin{minipage}[t]{0.47\hsize}
\centering
\includegraphics[scale = 0.56, clip]{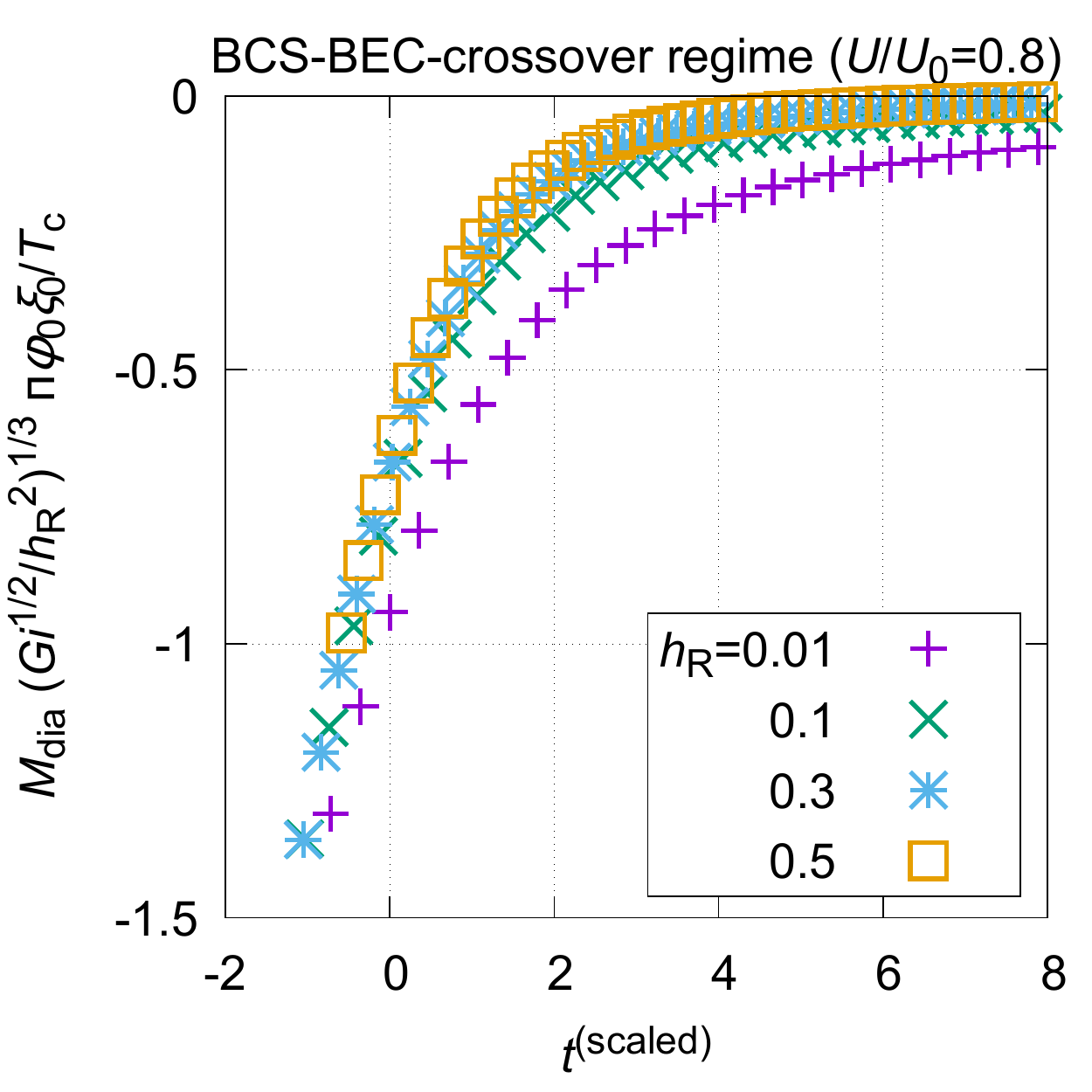}
\caption{
LLL-scaling plot of the SCF-induced magnetization in the BCS-BEC-crossover regime.
The symbols are used in the same way as in Fig.~\ref{Fig:Mdia_Scaled_U05}.
}
\label{Fig:Mdia_Scaled_U08}
\end{minipage}
\end{tabular}
\end{figure*}

\subsection{Specific heat and diamagnetic susceptibility}
\label{SubSec:Specific_heat_and_diamagnetic_susceptibility}

Before discussing the calculated specific heat and diamagnetic susceptibility, we mention the well-known results in the Gaussian approximation in the low-field limit.
We first consider the case where the mode coupling is so weak as in the BCS regime that $\xiR \simeq \xi_0$ and $\TcR \simeq \Tc$.
At temperatures higher than the critical temperature where the mode coupling between SCF is negligible (\textit{i.e.}, outside the critical region), it is known [see Eq.~(\ref{Eq:SpecificHeat_Gaussian})] that the SCF-induced specific heat $\widetilde{c}_V^\text{(Gauss)}$ is represented as
\begin{equation}
\widetilde{c}_V^\text{(Gauss)} = \frac{1}{8 \pi} \frac{1}{{\xi_0}^3 \sqrt{T / \Tc - 1}},
\label{Eq:cVGauss}
\end{equation}
and consistently that the SCF-induced diamagnetic susceptibility $\widetilde{\chi}_\text{dia}^\text{(Gauss)}$ is represented as~\cite{Schmid_1969}
\begin{equation}
\widetilde{\chi}_\text{dia}^\text{(Gauss)} = - \frac{\pi \mu_0}{6 {\phi_0}^2} \frac{\xi_0 \Tc}{\sqrt{T / \Tc - 1}},
\label{Eq:chidiaGauss}
\end{equation}
where $\mu_0$ is the vacuum permeability.
On the other hand, if the mode coupling is strong as in the BCS-BEC-crossover regime, the experimentally determined coherence length and critical temperature will be the renormalized ones $\xiR$ ($> \xi_0$) and $\TcR$ ($< \Tc$), respectively.
Thus, if we use these experimentally determined values, Eqs.~(\ref{Eq:cVGauss}) and (\ref{Eq:chidiaGauss}) should be replaced respectively with
\begin{equation}
c_V^\text{(Gauss)} = \frac{1}{8 \pi} \frac{1}{{\xiR}^3 \sqrt{T / \TcR - 1}}
\label{Eq:cVGaussR}
\end{equation}
and
\begin{equation}
\chi_\text{dia}^\text{(Gauss)} = - \frac{\pi \mu_0}{6 {\phi_0}^2} \frac{\xiR \TcR}{\sqrt{T / \TcR - 1}}.
\label{Eq:chidiaGaussR}
\end{equation}
Therefore, \textit{even the expression of thermodynamic quantities in the Gaussian approximation can be greatly affected by the mode coupling in the BCS-BEC-crossover regime}.

First, let us start with our numerical results of the specific heat in the low-field limit.
Figures~\ref{Fig:cV_U05} and \ref{Fig:cV_U08} respectively show the temperature dependence of the specific heat in the BCS and BCS-BEC-crossover regimes.
In each figure, the red solid line is our numerical result, while the black dotted line is the analytical result in the Gaussian approximation [Eq.~(\ref{Eq:cVGaussR})].

In the BCS regime (Fig.~\ref{Fig:cV_U05}), the numerical line is in accord with the result in the Gaussian approximation if $T$ is close to $\TcR$ ($T / \TcR - 1 \lesssim 0.01$), which is consistent with the fact that mode coupling is so weak (\textit{i.e.}, $Gi \sim 10^{-6} \ll 1$) that the critical behavior cannot appear except in a very narrow temperature range ($T / \TcR - 1 \lesssim 10^{-6}$).
When $T / \TcR - 1 \gtrsim 0.01$, on the other hand, the numerical value is smaller than the Gaussian-approximation value.
This is simply caused by the high-energy cutoff $c$ [see Eqs.~(\ref{Eq:CutoffCondition}) and (\ref{Eq:CutoffCondition_Mag})], which is not used in the conventional Gaussian approximation and effectively suppresses the SCF-induced thermodynamic response.

In the BCS-BEC-crossover regime (Fig.~\ref{Fig:cV_U08}), the numerical line does not fit anymore the result in the Gaussian approximation due to the strong mode coupling.
In contrast to the BCS regime (Fig.~\ref{Fig:cV_U05}), the numerical value is larger than the Gaussian-approximation value when the temperature is relatively far from $\TcR$ ($T / \TcR - 1 \gtrsim 0.03$).

Next, we show our numerical results of the diamagnetic susceptibility in the low-field limit.
Figures~\ref{Fig:chi_U05} and \ref{Fig:chi_U08} respectively show the temperature dependence of the diamagnetic susceptibility in the BCS and BCS-BEC-crossover regimes.
In each figure, the red solid line is the numerical result, while the black dotted line is the analytical result in the Gaussian approximation [Eq.~(\ref{Eq:chidiaGaussR})] in the same way as in Figs.~\ref{Fig:cV_U05} and \ref{Fig:cV_U08}.

In the BCS regime (Fig.~\ref{Fig:chi_U05}), the Gaussian approximation is appropriate as the specific-heat result if $T / \TcR - 1 \lesssim 0.01$ and the cutoff effect appears if $T / \TcR - 1 \gtrsim 0.01$.
We note that this kind of cutoff effect on diamagnetic susceptibility has been experimentally observed in conventional superconductors~\cite{Mosqueira_Carballeira_2001}.

In the BCS-BEC-crossover regime (Fig.~\ref{Fig:chi_U08}), the numerical value is larger than the Gaussian-approximation value, which is caused by the strong mode coupling as in the case of specific heat.
The obtained feature is qualitatively consistent with the data~\cite{Kasahara_Yamashita_2016} in FeSe.

\begin{figure*}[htbp]
\begin{tabular}{c}
\begin{minipage}[t]{0.47\hsize}
\centering
\includegraphics[scale = 0.56, clip]{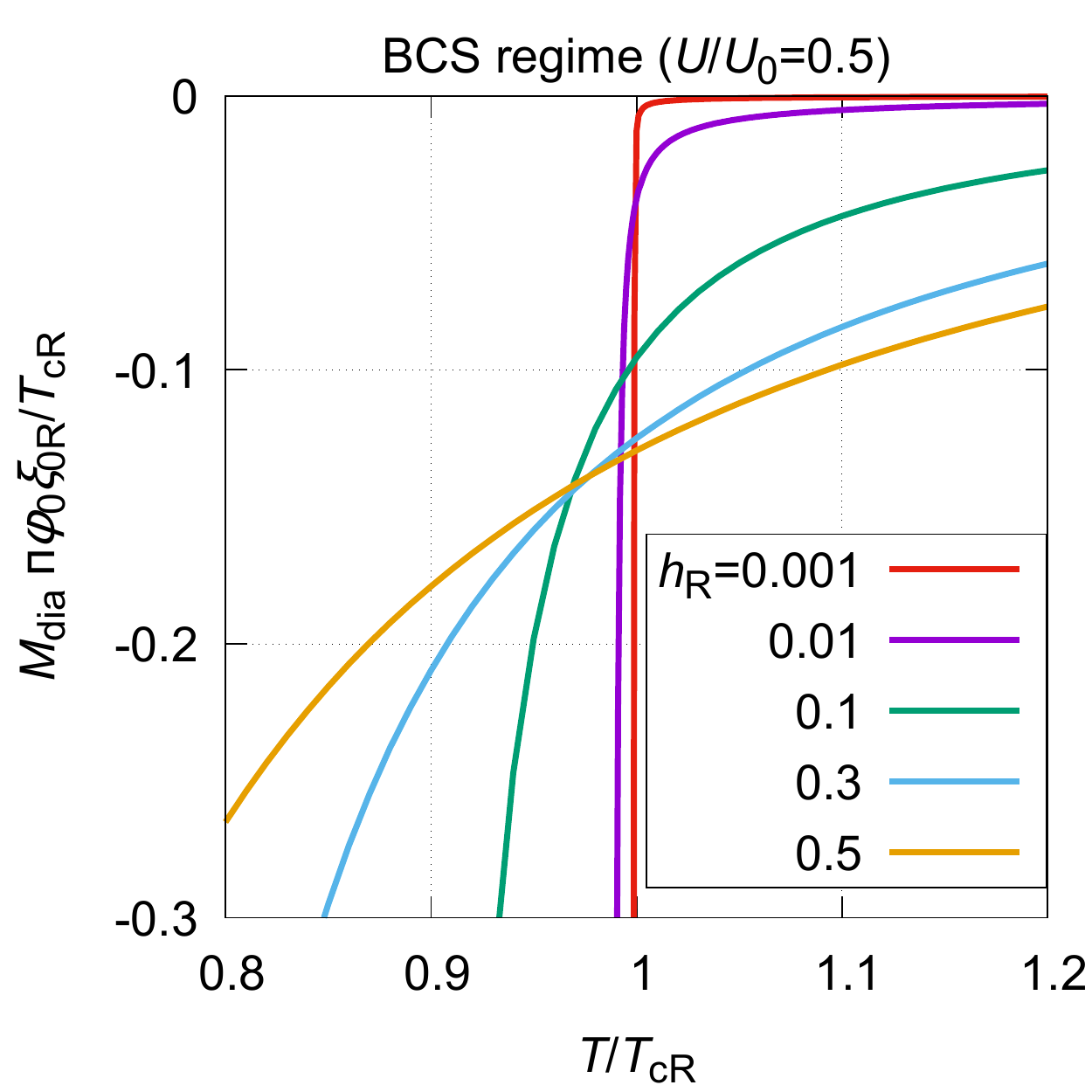}
\caption{
SCF-induced magnetization as a function of temperature in the BCS regime.
$h_\text{R}$ is a magnetic field in units of $\BctwoR (0)$.
We note that $\Mdia$ is measured in units of $(\pi \phi_0 \xiR / \TcR)^{-1}$.
}
\label{Fig:Mdia_U05}
\end{minipage}
\begin{minipage}[t]{0.03\hsize}
\hspace{0pt}
\end{minipage}
\begin{minipage}[t]{0.47\hsize}
\centering
\includegraphics[scale = 0.56, clip]{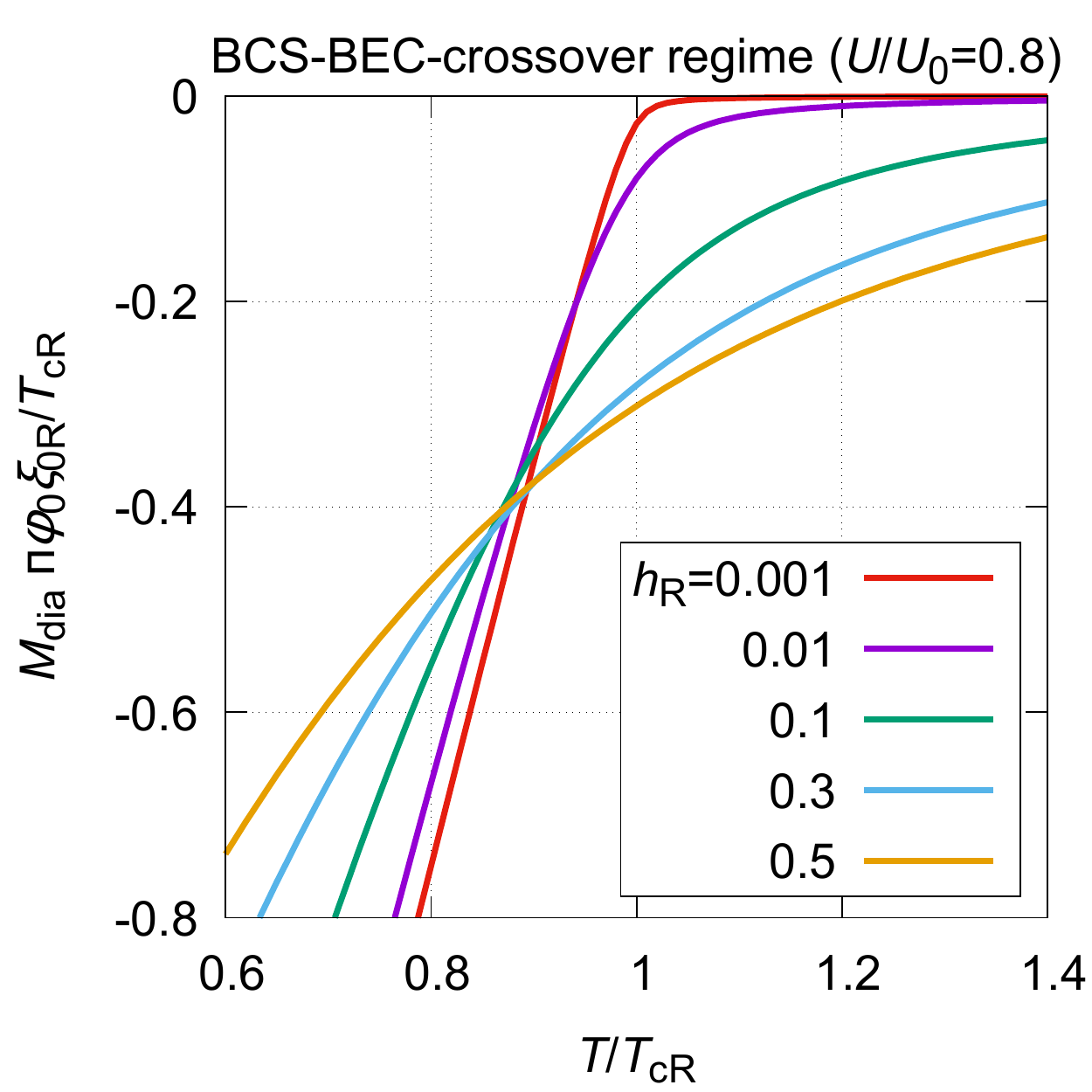}
\caption{
SCF-induced magnetization as a function of temperature in the BCS-BEC-crossover regime.
The symbols are used in the same way as in Fig.~\ref{Fig:Mdia_U05}.
We see from Figs.~\ref{Fig:Tc} and \ref{Fig:xi0R} that the unit of $\Mdia$, $(\pi \phi_0 \xiR / \TcR)^{-1}$, is about $100$ times larger than that in the BCS regime (Fig.~\ref{Fig:Mdia_U05}).
}
\label{Fig:Mdia_U08}
\end{minipage}
\end{tabular}
\end{figure*}

\subsection{Lowest-Landau-level scaling}
\label{SubSec:Lowest-Landau-level scaling}

In this subsection, we investigate whether the obtained specific heat and magnetization obey the so-called lowest-Landau-level (LLL) scaling around $\BctwoR (T)$, the renormalized $\Bctwo (T)$ [see Eq.~(\ref{Eq:Bc2R})].
Before moving on to the results, we review the properties of SCF in high magnetic fields, including the LLL scaling.
In high magnetic fields, if the mode coupling between SCF is moderately strong, the LLL ($N = 0$) modes of the order parameter field $\psi (\rbm)$ in Eq.~(\ref{Eq:OP_Mag}) have a dominant impact on thermodynamic and transport properties~\cite{Tesanovic_Xing_1992} compared to other higher-LL modes [$N \geq 1$ modes of $\psi (\rbm)$].
Such restriction of SCF to the LLL modes leads to the effective reduction of dimensionality (from 3D to 1D)~\cite{Ikeda_Ohmi_1989}, which changes the second-order $\Bctwo$ transition in the mean-field approximation to a crossover and creates the first-order vortex-melting transition~\cite{Nattermann_Scheidl_2000}.
Moreover, the restriction to the LLL modes simplifies dependences of the free energy on the temperature and the magnetic field~\cite{Ruggeri_Thouless_1976}, so that temperature and field dependences of thermodynamic quantities such as the specific heat and the magnetization are also simplified~\cite{Welp_Fleshler_1991} as follows:
\begin{eqnarray}
\frac{c_V}{\Delta c_V} &=& F_1 (t^\text{(scaled)}), 
\label{Eq:Scaling_cV}
\\
\Mdia \times \left( \frac{\sqrt{Gi}}{{\hR}^2} \right)^{1/3} &=& F_2 (t^\text{(scaled)}).
\label{Eq:Scaling_Mdia}
\end{eqnarray}
Here, $\Delta c_V$ is the mean-field specific-heat jump [see Eq.~(\ref{Eq:GinzburgNum})], $F_1 (x)$ and $F_2 (x)$ are some scaling functions, the dimensionless temperature $t^\text{(scaled)}$ is
\begin{equation}
t^\text{(scaled)} = \frac{\epsilonR + \hR}{(\sqrt{Gi} \, \hR)^{2/3}},
\label{Eq:Scaling_t}
\end{equation}
$\epsilonR$ ($= T / \TcR - 1$) is the dimensionless temperature, and $\hR$ [$= 2 \pi {\xiR}^2 B / \phi_0 \equiv B / B_\text{c2R}(0)$] is the dimensionless magnetic field, where $B_\text{c2R} (T)$ is the renormalized counterpart of $\Bctwo (T)$ [see Eq.~(\ref{Eq:Bc2R})].
The relations such as Eqs.~(\ref{Eq:Scaling_cV}) and (\ref{Eq:Scaling_Mdia}), which are based on the restriction of SCF to the LLL modes, are called the LLL scaling.
By using the LLL-scaling plot, we can check whether the LLL modes are dominant or not.

Figures~\ref{Fig:cV_Scaled_U05} and \ref{Fig:cV_Scaled_U08} show the LLL-scaling plots of the specific heat in the BCS and BCS-BEC-crossover regime, respectively.
From Fig.~\ref{Fig:cV_Scaled_U05}, we see that the LLL scaling is satisfied, which means that the LLL modes are dominant in the BCS regime.
On the other hand, Fig.~\ref{Fig:cV_Scaled_U08} shows the LLL scaling breaks down, which means that the higher-LL modes are important as well as the LLL modes in the BCS-BEC-crossover regime.
This can be understood as an effect of the strong mode coupling, which makes the contribution of the LLL modes less dominant.
In fact, the renormalized fluctuation theory for the weak-coupling BCS regime has indicated that the higher-LL modes are not negligible in describing the SCF in lower fields satisfying~\cite{Ikeda_1995}
\begin{equation}
B < \BctwoR (0) \, Gi.
\end{equation}
This expression implies that, with increasing the mode coupling, the LLL scaling should break down.

Figures~\ref{Fig:Mdia_Scaled_U05} and \ref{Fig:Mdia_Scaled_U08} show the LLL-scaling plots of the magnetization in the BCS and BCS-BEC-crossover regime, respectively.
The LLL scaling is satisfied in the BCS regime, while it breaks down in the BCS-BEC-crossover regime consistently with the result of the specific heat.

\subsection{Crossing of magnetization curves}
\label{SubSec:Crossing_of_magnetization_curves}

In this subsection, the crossing behavior of the magnetization curves, which has been experimentally observed in FeSe~\citep{Kasahara_Yamashita_2016}, is investigated in both the BCS and BCS-BEC-crossover regimes.
Figures~\ref{Fig:Mdia_U05} and \ref{Fig:Mdia_U08} show the obtained SCF-induced magnetization curves in the BCS and BCS-BEC-crossover regimes, respectively.
We note that the LLL-scaling plots shown in Figs.~\ref{Fig:Mdia_Scaled_U05} and \ref{Fig:Mdia_Scaled_U08} can be obtained respectively from Figs.~\ref{Fig:Mdia_U05} and \ref{Fig:Mdia_U08} by using $t^\text{(scaled)}$ instead of $T$ as the horizontal axis and $\Mdia \times (\sqrt{Gi} / {\hR}^2)^{1/3}$ instead of $\Mdia$ as the vertical axis.
The curves in Fig.~\ref{Fig:Mdia_U05} show a crossing behavior in the field range $0.1 \lesssim \hR \lesssim 0.5$.
On the other hand, the curves in Fig.~\ref{Fig:Mdia_U08} show a crossing in $0.01 \lesssim \hR \lesssim 0.5$.
Therefore, in the BCS-BEC-crossover regime, the field range where the crossing behavior appears is broad compared to that in the BCS regime.

These results of the LLL-scaling plot and the crossing behavior are summarized as follows: in an isotropic 3D system in the BCS regime, mainly the LLL modes ($N = 0$) of SCF create the crossing behavior only in a high field range ($0.1 \lesssim \hR \lesssim 0.5$).
On the other hand, in the BCS-BEC-crossover regime with stronger SCF, the higher-LL modes ($N \geq 1$) of SCF in addition to the LLL modes create the crossing behavior in a broader field range ($0.01 \lesssim \hR \lesssim 0.5$).

\section{Discussion}
\label{Sec:Discussion}

We discuss relevance of our results to the anomalous SCF-induced diamagnetic response observed in FeSe.
First, our numerical result in the BCS-BEC-crossover regime (Fig.~\ref{Fig:chi_U08}) is qualitatively consistent with the large diamagnetic susceptibility observed in FeSe~\cite{Kasahara_Yamashita_2016}.
Second, the experimentally observed crossing behavior of magnetization curves~\cite{Kasahara_Yamashita_2016} may also be explained based on our result in the BCS-BEC-crossover regime (Fig.~\ref{Fig:Mdia_U08}).
Although we have tried to quantitatively fit our numerical results to the experimental data, no quantitative agreement has been obtained.
This may be due to our neglect of the detailed band structure of FeSe in the present theory, which starts from an isotropic 3D continuum model.


FeSe is considered to be a two-band system consisting of hole and electron bands, and these two bands are asymmetric.
That is, roughly speaking, the hole and electron bands seem to be in the BCS and BCS-BEC-crossover regimes, respectively~\cite{Kasahara_Watashige_2014}.
In our previous work~\cite{Adachi_Ikeda_2016}, we have studied the SCF-induced diamagnetic response in a symmetric two-band system in the BCS regime and shown that the diamagnetic susceptibility can become larger due to high-energy modes of SCF than that in a single-band system.
This can lead to an expectation that the diamagnetic susceptibility in a symmetric two-band system in the BCS-BEC-crossover regime is also enhanced further compared with that in a single-band system explored in the present study.
However, we cannot directly apply the scheme used in the present study to a more general asymmetric two-band system such as FeSe since the gradient expansion in the GL action might not be justified~\cite{Koshelev_Varlamov_2014} due to the difference in the coherence length between the two bands.
Therefore, in considering a general two-band system to describe FeSe in detail, it would be better to treat the full momentum dependence of the SCF contribution by using a more microscopic method such as the self-consistent T-matrix approximation~\cite{Yanase_Yamada_1999}.

Before ending this section, we add some remarks.
We have shown in our previous~\cite{Adachi_Ikeda_2016} and the present works that, as observed in the anisotropic 3D system FeSe with $\xi_{0, c} > s / \sqrt{2}$ ($s$ is the interlayer spacing), appearance of the crossing behavior of the magnetization curves $\Mdia (T)$ over some field range is not limited to 2D-like systems such as a lot of high-$\Tc$ cuprates.
Therefore, contrary to an argument given elsewhere~\cite{Chi_Jiang_2017}, this crossing behavior has nothing to do with the dimensional crossover present in quasi-2D systems with $\xi_{0, c} < s / \sqrt{2}$.

\section{Conclusion}
\label{Sec:Conclusion}

In this paper, we have studied SCF effects on thermodynamic properties in the BCS-BEC-crossover regime by using a simple 3D electron model.
As a consequence, we conclude that the following three features can emerge due to the strong mode coupling between SCF, which is characteristic of electron systems in the BCS-BEC-crossover regime.
First, the SCF-induced specific heat and diamagnetic susceptibility can seemingly exceed the corresponding values in the Gaussian approximation (Figs.~\ref{Fig:cV_U08} and \ref{Fig:chi_U08}).
Second, the LLL scaling can break down (Figs.~\ref{Fig:cV_Scaled_U08} and \ref{Fig:Mdia_Scaled_U08}), which means that the higher-LL modes ($N \geq 1$) of SCF are important in addition to the LLL modes ($N = 0$).
Third, the crossing behavior of magnetization curves can appear in a broad range of magnetic fields (Fig.~\ref{Fig:Mdia_U08}), which is caused by both the LLL and higher-LL modes.

\section*{Acknowledgment}
\label{Sec:Acknowledgment}

We are grateful to Y. Matsuda and S. Kasahara for informative discussions.
The present research was supported by JSPS KAKENHI [Grants No. 16K05444 and No. 17J03883].
One of the authors (K. A.) thanks JSPS for support from a Research Fellowship for Young Scientists.


\bibliographystyle{apsrev4-1}
\bibliography{BCS-BEC}


\end{document}